\documentclass[11pt]{article}

\usepackage[margin=0.9in]{geometry}
\usepackage{caption} 
\usepackage{graphicx,psfrag,epsf}
\usepackage{xcolor}

\usepackage{enumerate}
\usepackage{apacite}
\usepackage{natbib}
\usepackage{url}
\usepackage{amsmath,latexsym,amssymb,amsthm,bm}
\newcommand{\blind}{1}
\usepackage{epigraph,xcolor}
\usepackage{graphicx}
\usepackage{multirow}
\usepackage{hyperref}
\hypersetup{colorlinks=true,citecolor=blue,
	pdfpagemode=FullScreen,linktocpage=true}

\usepackage{cleveref}

\newtheoremstyle{general}
{3mm} 
{3mm} 
{\it} 
{} 
{\bfseries} 
{.} 
{.5em} 
{} 

\def\N{\mathbb{N}}
\def\Z{\mathbb{Z}}

\def\E{\mathbb{E}}
\newcommand{\x}{\bm{X}}

\newcommand{\abs}[1]{\left\lvert #1 \right\rvert}
\newcommand{\norm}[1]{\left\| #1 \right\|}

\newcommand{\F}{\mathcal{F}}
\newcommand{\G}{\mathcal{G}}

\newcommand{\eps}{\varepsilon}
\renewcommand{\L}{\mathcal{L}}
\renewcommand{\N}{\mathcal{N}}

\newcommand{\given}{\hspace{2pt}\vert\hspace{2pt}}

\newcommand{\dash}{^{\prime}}
\newcommand{\ddash}{^{\prime\prime}}

\newcommand{\ind}{\mathbb{I}}

\theoremstyle{general}

\newtheorem{theorem}{Theorem}
\newtheorem{corollary}{Corollary}

\newtheorem{assumption}{Assumption}

\makeatletter
\renewenvironment{proof}[1][\proofname]{\par
    \pushQED{\qed}%
    \normalfont \topsep6\p@\@plus6\p@\relax
    \trivlist
    \item\relax{
        \bfseries
        #1\@addpunct{.}}\hspace\labelsep\ignorespaces
    }{%
     \popQED\endtrivlist\@endpefalse
     }
\makeatother

\numberwithin{equation}{section}
\numberwithin{theorem}{section}
\numberwithin{lemma}{section}
\numberwithin{defn}{section}
\numberwithin{corollary}{section}

\begin{document}
	
\date{}

\if 1\blind
{
\title{Nonparametric quantile regression for time series with replicated observations and its application to climate data}
\author{Soudeep Deb\thanks{Email: soudeep@iimb.ac.in. ORCiD: 0000-0003-0567-7339.} \\ Indian Institute of Management Bangalore \\ Bannerghatta Main Rd, Bangalore, KA 560076, India. \\ and  \\ Kaushik Jana\thanks{Email: kaushikjana11@gmail.com. ORCiD: 0000-0003-4832-1375} \\ Imperial College London\\Department of Mathematics,\\ 180 Queen's Gate, London SW7 2AZ, United Kingdom }
\maketitle
} \fi

\bigskip




\begin{abstract}
This paper proposes a model-free nonparametric estimator of conditional quantile of a time series regression model where the covariate vector is repeated many times for different values of the response. This type of data is abound in climate studies. To tackle such problems, our proposed method exploits the replicated nature of the data and improves on restrictive linear model structure of conventional quantile regression. Relevant asymptotic theory for the nonparametric estimators of the mean and variance function of the model are derived under a very general framework. We provide a detailed simulation study which clearly demonstrates the gain in efficiency of the proposed method over other benchmark models, especially when the true data generating process entails nonlinear mean function and heteroskedastic pattern with time dependent covariates. The predictive accuracy of the non-parametric method is remarkably high compared to other methods when attention is on the higher quantiles of the variable of interest. Usefulness of the proposed method is then illustrated with two climatological applications, one with a well-known tropical cyclone wind-speed data and the other with an air pollution data.  
\end{abstract}


\newpage

\section{Introduction}
\label{sec:introduction}

Replicated data is characterized by many repetitions of the covariate values at different values of the response variable. This is generally denoted by paired data of the form $(\x_t, Y_{tj})$, where $j=1,\ldots, n_t$ correspond to different values of the response variable for the $t^{th}$ realization of the covariate vector, i.e.\ $\x_t$ is replicated $n_t$ times in the dataset. Such data are frequently encountered in many contexts, including climate and environmental science study. For instance, the intensities of tropical cyclones (quantified by maximum wind speed during a particular cyclone) over a period of several years, with multiple observations within a year, have been studied in relation to the year of occurrence and other year-specific climate variables; see for example \cite{elsner2008increasing}, \cite{jagger2009modeling} and \cite{Kossin_2013}. Time series modelling of daily or sub-daily extreme precipitation levels were related to monthly El Nino Southern Oscillation (ENSO) index values in a few other studies, such as \cite{Nicholls1993} and \cite{Ropelewski2008}. Similar problems have been studied in financial time series as well, e.g.\ \cite{Embrechts1997} analyzed stock returns above a certain threshold in a given time period (say in a year), using the time period as a covariate. On the other hand, in public health growth research, it is common for scientists to study the dependence of certain quantiles of body dimensions on the age, ethnicity, gender etc.\ of the subject, by using cross-sectional data collected on birthdays of subjects (\cite{Redden_2004, Fernandez_2004}). It goes without saying that such type of data renders a good opportunity to study the quantiles of the response variable, which is in fact of paramount interest in many environmental problems. The reader is referred to \cite{huang2017quantile}, \cite{rivas2020trends} and \cite{vasseur2021comparing} for some relevant readings.

For a conventional set up of response-covariate space with the interest in specified quantiles, the linear quantile regression estimator of \cite{koenker1978regression} can be used for estimating the parameters and related inferences. In the case of replication in the covariate, \cite{Knight} and \cite{jana2019improving} used an alternative linear quantile regression model for the response sample quantiles at distinct covariate values and studied asymptotic behavior of regression estimates for replicated data. A few more similar applications can be found in \cite{lu2015weighted}, \cite{chen2021quantile} and the references therein. Intriguingly, we find that the existing literature always assumes a linear structure in the context of replicated data, albeit it has been well studied that climate data can exhibit highly nonlinear pattern (\cite{Watson-Parris2021}). While it is true that the paucity of the data and only a few distinct values of the covariates can pose a challenge to go beyond linear model, replications offer an option to do better. Furthermore, the time indexing of the variables provide an additional feature of imposing a dependence structure in the modelling. In this paper, we incorporate these features of the data and propose a nonparametric method of estimating conditional quantiles at different levels under a time-dependent framework. 

It is of the essence here to give a brief overview of nonparametric techniques in relation to the subject matter of this article. Literature on nonparametric methods in estimating the regression function in a standard setting dates back to more than three decades. The books by \cite{wahba1990spline} and \cite{fan2018local} are great resources to understand the introductory theory, while some applications of nonparametric methods in various contexts of climatology can be found in the works by \cite{lanzante1996resistant}, \cite{henry2002locating} and \cite{bercu2019nonparametric}. An attractive advantage of nonparametric approach is that it lets the data to ``speak for themselves'' and thereby enables one to detect the underlying regression structures in a better way. Naturally, nonparametric methods in the quantile regression problems have also been studied extensively, and several researchers have shown its effectiveness in identifying potentially nonlinear nature of the regression functions in the data. See \cite{leider2012quantile}, \cite{wei2019applications} and \cite{li2020nonparametric} for some interesting applications of it. These studies, however, do not take into account that the observations can be dependent among themselves, which is often the case for a real-life climate data. To that end, \cite{honda2000nonparametric}, \cite{cai2002regression} and \cite{ziegelmann2005nonparametric} are some of the early papers that derived results on the properties of the quantile estimates of the deterministic component of the underlying process under strong mixing conditions. \cite{dabo2012nonparametric}, \cite{honda2013nonparametric} and \cite{gregory2018smooth} are few other crucial improvements in this regard. Interestingly, to the best of our knowledge, the setting of a time series quantile regression with replicated observations has not been considered before. Not only we aim to bridge that gap in this article but we also consider a very general setup that would allow one to use the methods in any such problem. After describing the theoretical properties, we illustrate the usage of the proposed approach through a detailed simulation study as well as two different real data examples related to the climate studies.

Outline of the paper is as follows. The next section describes the notations, main methods and theoretical results. \Cref{sec:simulation} presents a simulation study under various setups and we demonstrate the advantages of the nonparametric approach over the existing methods for replicated data. Application and comparison of different methods on two real data sets are illustrated in \Cref{sec:application}. We conclude the main paper with some important remarks in \Cref{sec:conclusion}. Theoretical proofs of the theorems are presented in \Cref{sec:proof} while some additional simulation results are given in \Cref{sec:additional-results}.

\section{Methodology}
\label{sec:methods}

\subsection{Notations and existing results}
\label{subsec:notations}

In the following discussion, unless otherwise specified, $\overset{\mathcal{L}}{\rightarrow}$ denotes convergence in law, $\N(\eta,\Psi)$ stands for a normal distribution with mean parameter $\eta$ and dispersion parameter $\Psi$, $\abs{S}$ denotes the cardinality of a set $S$, and $\norm{\cdot}$ refers to the $\L_2$ norm. Recall that the $\L_p$ norm for the $q$-dimensional vector $w=(w_1,w_2,\hdots,w_q)^T$ is defined as
\begin{equation}
    \label{eqn:vector-norm}
    \norm{w}_p = \left(\sum_{i=1}^q \abs{w_i}^p\right)^{1/p}.
\end{equation}

Let $\mathbb{N}, \Z$ denote the set of natural numbers and the set of integers respectively. Throughout this study, we consider time-dependent data of the form $(\x_t,Y_{tj})_{1\leqslant t \leqslant n,j \in \Gamma_t}$, where $\Gamma_t$ denotes the index set (possibly unobserved) for the replicated observations at time $t$. Following is the regime we shall consider for developing the asymptotic theory in this work.

\begin{assumption}\label{asmp:regime}
The index set $\Gamma_t$ at time $t$, for $1\leqslant t\leqslant n$, satisfies the condition $\abs{\Gamma_t}\to\infty$ as $n\to\infty$.
\end{assumption}

In other words, we assume that there are large number of replications at every time point as the size of the data grows. It is also assumed that for a given covariate profile $\x_t$, $Y_{tj}$'s are conditionally independent with the common distribution $F_t$. For $\tau \in (0,1)$, the $\tau$-quantile of $F_t$ can be written as $q(\tau \given \x_t) = \inf\{y : P(Y_{tj} \leqslant y \given \x_t) \geqslant \tau \}$. The corresponding sample $\tau$-quantile of the observations is then defined as  
\begin{equation}
\label{eq:sample-quantile}
Q_t^{(\tau)} := \widehat  q(\tau \given \x_t) = \mathop{\arg\min}_{m_t} \sum_{j\in\Gamma_t} \rho_{\tau}(Y_{tj}-m_t).
\end{equation}
 where $\rho_{\tau}(u)=u(\tau-I(u<0))$.
 
Consider the following quantile regression model \citep{koenker2005} at probability level $\tau \in (0,1)$   
\begin{equation}
\label{eq:sqrm}
Q_t^{(\tau)} = \mu(\x_t) + \sigma(\x_t)F^{-1}(\tau),
\end{equation}
where $\mu(\cdot)$ and $\sigma^2(\cdot)$ are the mean function and the variance function, $F(\cdot)$ is an appropriately chosen error distribution, and $F^{-1}(\tau)$ denotes the quantile of it at level $\tau$. In particular, when $\mu(\cdot)$ is a linear function and $\sigma(\cdot)$ is constant, it is similar to the standard quantile regression model, hereafter abbreviated as SQRM. Let $\bm\beta_\tau$ be the coefficient vector for SQRM such that $\mu(\x_t)=\x_t'\bm\beta_\tau$, $\tau \in (0, 1)$, then $\bm\beta_\tau$ can be estimated as (\cite{koenker1978regression})
\begin{equation}
\label{eq:estimate-kb}
	\widehat {\bm\beta}_{\tau;kb} = \underset{\bm\beta}{\arg\min}\sum_{t=1}^n\sum_{j\in \Gamma_t}
	\rho_\tau(Y_{tj}- \x'_t \bm\beta_\tau).
\end{equation}

Detailed discussion on this can be found in \cite{koenker2005}. In a lot of studies, SQRM has been used effectively to analyze climate data, see \cite{jagger2009modeling}, \cite{ying2011climate} for example. 
Note that this benchmark method can be naturally applied under the setup of replicated observations. 

In the framework of replicated data, \cite{jana2019improving} improved upon the above estimate in an asymptotically efficient sense. They used the asymptotic distribution for $Q_t^{(\tau)}$ to propose the following weighted least squares estimate for $\bm\beta_\tau$:
\begin{equation}
\label{eq:estimate-js}
	\widehat {\bm\beta}_{\tau;js}= \left(\x'\hat\Omega_\tau^{-1}\x\right)^{-1}\x'\hat\Omega_\tau^{-1}\bm{Q}^{(\tau)}.
\end{equation}

In the above expression, $\x = [\x_1:\hdots:\x_n]'$, $\bm{Q}^{(\tau)} = (Q_1^{(\tau)},\hdots,Q_n^{(\tau)})'$ and $\hat\Omega_\tau$ is a consistent estimate of the diagonal matrix $\Omega_\tau$ with $t^{th}$ diagonal element as 
\begin{equation*}
    (\Omega_\tau)_{t,t} = \frac{\tau(1-\tau)}{\abs{\Gamma_t} f_t^2(F_t^{-1}(\tau))},
\end{equation*}
where $f_t(\cdot)$ is the density of $F_t(\cdot)$, the error distribution associated with the $t^{th}$ observation.

\begin{theorem}[Due to \cite{jana2019improving}]
\label{thm:jana}
Let $N=\sum \abs{\Gamma_t}$ and assume that as $N \to \infty$ the vector $(\abs{\Gamma_t}/N)_{1\leqslant t\leqslant n}$ converges in Euclidean norm to a vector $(\xi_t)_{1\leqslant t\leqslant n}$ with positive components. $F_t$, for $1\leqslant t\leqslant n$, are considered to be absolutely continuous, with continuous density function $f_t$ uniformly bounded away from $0$ and $\infty$ at  $F^{-1}_t(\tau)$. Let $\max_t\norm{\x_t} = o(\sqrt{N})$ and assume that the sample matrices
$D_{jn}=N^{-1}\!\sum_{t=1}^n \abs{\Gamma_t}\ \{f_t(F^{-1}_t(\tau))\}^j {\x}_t {\x}_t'$, for $j=0,1,2$, converge to positive definite matrices $D_j$, as $N\rightarrow\infty$. Then, the following results hold.
\begin{itemize}
    \item[(a)] $\sqrt N(\hat{\bm\beta}_{\tau;kb} - \bm\beta_\tau) \overset{\mathcal{L}}{\rightarrow} \N\left(\bm0,\tau(1-\tau)D_1^{-1}D_0D_1^{-1}\right).$
    \item[(b)] $\sqrt N(\hat{\bm\beta}_{\tau;js} - \bm\beta_\tau) \overset{\mathcal{L}}{\rightarrow} \N\left(\bm0,\tau(1-\tau)D_2^{-1}\right).$
    \item[(c)] The limiting dispersion matrix of $\sqrt N(\hat{\bm\beta}_{\tau;kb} - \bm\beta_\tau)$ is larger than or equal to that of $\sqrt N(\hat{\bm\beta}_{\tau;js} - \bm\beta_\tau)$ in the sense of L\"{o}wner order.
\end{itemize}
\end{theorem}

Proof of the above theorem is available on the aforementioned paper and is omitted from here for brevity. The result establishes the superiority of the updated method over the conventional method in an asymptotic sense. The authors also presented detailed simulation study and an application to climate data to demonstrate that $\hat{\bm\beta}_{\tau;js}$ serves as a better estimator for the higher quantiles. Later in this paper, we shall compare the performance of our proposed method against these two benchmark approaches (quantile analysis through $\hat{\bm\beta}_{\tau;kb}$ and $\hat{\bm\beta}_{\tau;js}$) and will refer to them as KB and JS methods, respectively.

\subsection{Main assumptions and proposed method}
\label{subsec:main-method}

As we pointed out in the introduction, the aforementioned standard techniques fall short when $\mu(\cdot)$ and $\sigma(\cdot)$ are expected to deviate from the assumptions of SQRM. Especially, $\mu(\cdot)$ is often nonlinear in case of quantile regression models in climate data, see for example \cite{ding2017influence}. Furthermore, either the structure of replicated data or the temporal dependence is not utilized in the benchmark methods. In our attempt to develop an alternate methodology on that front, we rely on the fact that sample quantiles converge in probability to true quantiles under \Cref{asmp:regime}. Then, for a fixed $\tau\in (0,1)$, we consider the stochastic regression model
\begin{equation}
\label{eq:main-model}
Q_t^{(\tau)} = \mu(\x_t) + \sigma(\x_t)e_t, \qquad t=1,2,\hdots,n,
\end{equation}
where $e_t$ are independent and identically distributed (iid) random errors with $\E(e_t)=0$ and $\E(e_t^2)=1$. We keep similar notations as before to avoid confusion. Here, $\mu(\cdot)$ and $\sigma^2(\cdot)$ correspond to respectively the mean function and the conditional variance function of the model. $\sigma(\cdot)$ is considered to be strictly positive. Both functions are unknown and need to be estimated. Note that \cref{eq:main-model} is general enough to include the most common linear and nonlinear regression models for the quantiles. We propose to use nonparametric techniques to develop an improvised version of $\mu(\cdot)$ and $\sigma(\cdot)$ in regressing quantiles for replicated data and that is the primary contribution of this work. To that end, we consider a very general class of stationary processes as functions of iid random variables for the regressor $\x_t$. Following are the relevant assumptions.

\begin{assumption}
\label{asmp:regressor-process}
Let $\eps_j, j\in\Z,$ be iid random variables. Then,
\begin{equation}
\label{eqn:regressor-process}
\x_i = h(\eps_{i-s}; s \in \Z), \qquad i\in\Z,
\end{equation}
where $h$ is a measurable function such that $\x_i$ is well-defined.
\end{assumption}

\begin{assumption}
\label{asmp:independence}
$e_t$ in \cref{eq:main-model} is independent of $\F_t$, the $\sigma$-field generated by $(\hdots,\eps_{t-1},\eps_t)$.
\end{assumption}

Following the coupling idea of \cite{wu2005nonlinear}, we next define a dependence measure for $\x_t$. Let $(\eps_i\dash)_{i\in\Z}$ be an iid copy of $(\eps_i)_{i\in\Z}$, and let $\eps_i^* = \eps_i$ if $i\ne 0$ and $\eps_0^* = \eps_0^{\prime}$. We use $\F^*_t$ to denote the sigma-field generated by $(\hdots,\eps_{t-1}^*,\eps_t^*)$. Let $G_k(\cdot\given\F_t)$ be the conditional distribution function of $\x_{t+k}$ given $\F_t$ and $g_k(\cdot\given\F_t)$ be the associated conditional density. $G_k(\cdot\given\F_t^*)$ and $g_k(\cdot\given\F_t^*)$ are defined similarly. Then, as \cite{wu2005nonlinear} pointed out, 
\begin{equation}
\label{eq:dependence-measure}
    \theta_t = \sup_{x} \norm{g_t(x\given\F_0) - g_t(x\given\F_0^*)} 
\end{equation}
signifies the contribution of $\eps_0$ in predicting $\x_{t}$. Further, let
\begin{equation}
\label{eq:sum-dependence-measure}
    \Theta_n = \sum_{t=1}^n \theta_t, \; \Theta_\infty = \sum_{t=1}^\infty \theta_t.
\end{equation}

\begin{assumption}
\label{asmp:srd}
$\Theta_\infty < \infty$, which corresponds to short-range dependent processes.
\end{assumption}

It is clear that the above setup provides a general and attractive framework. In the following discussion, wherever appropriate, $G(\cdot)$ and $g(\cdot)$ would denote the distribution function and the corresponding density function of $\x_t$. We estimate $g(\cdot)$ using the kernel density form
\begin{equation}
\label{eq:kernel-density}
\hat g(x; b_n) = \frac{1}{nb_n}\sum_{t=1}^n K\left(\frac{x-\x_t}{b_n}\right),
\end{equation}
where $b_n$ is a bandwidth sequence such that $b_n \to 0, nb_n\to\infty$ for $n\to\infty$; and $K(\cdot)$ is a kernel function whose properties are given by the following.

\begin{assumption}
\label{asmp:kernel}
For nonparametric estimation purposes, the kernel functions $K(\cdot)$ used in this study are assumed to be bounded, symmetric, Lipschitz continuous, have bounded derivative and satisfy the conditions $\int K(u)\ du=1$, $\phi_K = \int K^2(u)\ du < \infty$ and $\psi_K = \int u^2K(u)\ du < \infty$.
\end{assumption}

Finally, in developing the asymptotic theory, we shall require some regularity conditions on the behavior of the mean function, the variance function and the above-mentioned density functions.

\begin{assumption}
\label{asmp:regularity-conditions}
For a fixed $x$ and for $\epsilon > 0$, denote the $\epsilon$-ball of $x$ by $B(x,\epsilon)$. We call $x$ to be a favourable point if for some $\epsilon > 0$, $\mu(\cdot)$, $\sigma(\cdot)$ and $g(\cdot)$ have bounded fourth-order derivatives in $B(x,\epsilon)$, $\inf_{B(x,\epsilon)}\sigma(x) > 0$, $\inf_{B(x,\epsilon)}g(x) > 0$ and $\sup_{B(x,\epsilon)}g_1(x\given\F_0) < M$ for some constant $M$.
\end{assumption}

We now move on to estimating our main model. Since the exact expression of $\mu(\cdot)$ is assumed to be unknown, for the data $(\x_t,Y_{tj})_{1\leqslant t \leqslant n,j \in \Gamma_t}$, we are going to use the Nadaraya-Watson type estimator defined as 
\begin{equation}
\label{eq:mu-estimate}
\hat\mu(x; b_n) = \frac{1}{nb_n \hat g(x; b_n)}\sum_{t=1}^n K\left(\frac{x-\x_t}{b_n}\right)\mathop{\arg\min}_{m_t} \sum_{j\in\Gamma_t} \rho_{\tau}(Y_{tj}-m_t).
\end{equation}

The above expression gives us a point estimate of $\mu(\cdot)$. In the theorem below, we state a crucial large sample property of the estimate. On one hand, the theorem provides information regarding the asymptotic bias and consistency of the estimate while on the other, it can be used to construct confidence interval for the mean function. Detailed proof of the theorem is relegated to \Cref{sec:proof}. 

\begin{theorem}
\label{thm:clt}
Let $x$ be a favourable point in the sense of \Cref{asmp:regularity-conditions}. Then, under Assumptions \ref{asmp:regime}, \ref{asmp:regressor-process}, \ref{asmp:independence}, \ref{asmp:srd} and \ref{asmp:kernel}, as $n\to\infty$,
\begin{equation}
\label{eq:clt-mu}
\sqrt{nb_n}\left(\frac{\sqrt{\hat g(x;b_n)}}{\sigma(x)\sqrt{\phi_K}}\right)  \biggl( \hat\mu(x; b_n) - \mu(x) - \delta(x;b_n) \biggr) \overset{\mathcal{L}}{\rightarrow} \N(0,1),
\end{equation}
where $\delta(x;b_n) = b_n^2\psi_K(\mu\ddash(x) + 2\mu\dash(x)g\dash(x)/g(x))$ is the asymptotic bias of the estimate.
\end{theorem}

One can use the theorem directly to compute the confidence interval for $\mu(x)$. However, that requires either a consistent way of estimating $\delta(x;b_n)$ or more assumptions on $\mu\dash(x)$ and $\mu\ddash(x)$. In order to avoid that, we follow a jackknife type bias correction procedure, similar to \cite{wu2007inference}. The following corollaries to the above theorem provide us the required result on that note.

\begin{corollary}
\label{corr:jackknife}
For $\lambda>1$, let $\hat\mu^*_\lambda(\cdot;\cdot)$ be defined as 
\begin{equation}
\label{eq:jackknife-estimate}
    \hat\mu^*_\lambda(x;b_n) = \frac{\lambda\hat\mu(x;b_n) - \hat\mu(x;\sqrt\lambda b_n)}{\lambda - 1}.
\end{equation}
If $K_\lambda(\cdot)$ is the kernel function satisfying $K_\lambda(u)=(\lambda K(u)-\lambda^{-1/2}K(u/\sqrt\lambda))/(\lambda-1)$, then
\begin{equation}
\label{eq:clt-jackknife}
\sqrt{nb_n}\left(\frac{\sqrt{\hat g(x;b_n)}}{\sigma(x)\sqrt{\phi_{K_\lambda}}}\right)  \biggl( \hat\mu^*_\lambda(x; b_n) - \mu(x)\biggr) \overset{\mathcal{L}}{\rightarrow} \N(0,1),
\end{equation}
\end{corollary}

\begin{corollary}
\label{corr:ci-mu}
Let $\hat\sigma(x)$ be a consistent estimate of $\sigma(x)$. Then, a $100(1-\alpha)\%$ confidence interval for $\mu(x)$ is
\begin{equation}
\label{eq:ci-mu}
\hat\mu^*_\lambda(x;b_n) \pm z_{\alpha/2}\left(\frac{\hat\sigma(x)\sqrt{\phi_{K_\lambda}}}{\sqrt{nb_n \hat g(x;b_n)}}\right).
\end{equation}
\end{corollary}

Both corollaries follow directly from \Cref{thm:clt}, and the proofs are trivial. While the above results are true for any $\lambda>1$, experimentation suggested that $\lambda=2$ works as good as any other value and this choice also makes the computation convenient for us. Thus, in all applications below, we would use $\hat\mu^*_2(x;b_n)$. For notational simplicity, we are going to drop the subscript and write it as $\hat\mu^*(x;b_n)$.

Next, in order to implement the above result, we have to find a consistent estimate $\hat\sigma(x)$. Observe that $Q_t^{(\tau)}-\hat\mu^*(\x_t;b_n)$ is the residual from the model. Thus, a natural estimator of $\sigma^2(x)$ is
\begin{equation}
\label{eq:sigma-estimate}
    \hat\sigma^2(x;b_n) = \frac{1}{nb_n\hat g(x;b_n)}\sum_{t=1}^n K_2\left(\frac{x-\x_t}{b_n}\right) \left(Q_t^{(\tau)}-\hat\mu^*(\x_t;b_n)\right)^2.
\end{equation}

\begin{theorem}
\label{thm:sigma-consistency}
Let $x$ be a favourable point in the sense of \Cref{asmp:regularity-conditions} and let assumptions \ref{asmp:regime}, \ref{asmp:regressor-process}, \ref{asmp:independence}, \ref{asmp:srd} and \ref{asmp:kernel} to be true. Further assume that $\log n/nb_n^3 \to 0$. Then, $\hat\sigma^2(x;b_n) \to \sigma^2(x)$ in probability as $n\to\infty$.
\end{theorem}

Proof of this theorem is deferred to \Cref{sec:proof}. In the next subsection, we describe the implementation techniques of the method and the evaluation criteria to be used in the applications.

\subsection{Implementation}
\label{subsec:implementation}

We have established crucial asymptotic results of the nonparametric estimates for $\mu(x)$ and $\sigma^2(x)$. In practice, in order to implement the proposed method, we need to choose $b_n$ and $K(\cdot)$ appropriately. 

Recall that, in the calculation of $\hat g(x;b_n)$ and $\hat\mu(x;b_n)$, the bandwidth sequence needs to satisfy the conditions $b_n \to 0, nb_n\to\infty$ as $n\to\infty$. We also note the additional restriction on the bandwidth sequence in \Cref{thm:sigma-consistency} for ensuring consistency of $\hat\sigma^2(x;b_n)$. It is easy to observe that all these conditions are actually satisfied if we take $b_n=n^{-1/5}$, one of the most common choices in similar nonparametric studies. Thus, we propose to use this choice in all applications. As we shall see in the following sections, it provides good results in all setups. That being said, one can also take a cross-validation type approach to select the bandwidth sequence in a more objective way. Additionally, we want to point out that the estimates in \cref{eq:mu-estimate} and \cref{eq:sigma-estimate} can be potentially computed with different choices of bandwidth sequences. As long as the relevant assumptions are true, the asymptotic results hold.

So far as the kernel function is considered, it has been established in many literature that the estimates are not sensitive to the choice of the kernel. In all our computations, we use the kernel function $K(u)=(3/\pi)\max\{0,(1 - \norm{u}^2)\}^2$, which satisfies \Cref{asmp:kernel}.

In the remaining sections of the paper, we assess the efficacy of the proposed nonparametric method (NP) along with the other two approaches (KB and JS) mentioned before. Throughout, we compare the performances of the three different estimators of $\mu(x)$, calculated via $x'\widehat {\bm\beta}_{\tau;kb}$, $x'\widehat {\bm\beta}_{\tau;js}$, and $\hat {\mu}(x, b_n)$, as defined in \cref{eq:estimate-kb}, \cref{eq:estimate-js}, and \cref{eq:mu-estimate}, respectively. Primarily, the comparison is done through the root mean squared error (RMSE) criteria. Let $\mathcal{T}_1$ (respectively $\mathcal{T}_2$) denote the training set (respectively the test set) and suppose $\hat\mu(x)$ is the fitted value (respectively the prediction) from a model. Then, the training RMSE is calculated as follows:

\begin{equation}
    \label{eq:training-RMSE}
    \mathrm{Training \; RMSE} = \sqrt{\frac{1}{\abs{\mathcal{T}_1}}\sum_{x\in\mathcal{T}_1}(\hat\mu(x)-\mu(x))^2}.
\end{equation}

For evaluating the prediction accuracy, along with the RMSE, we compute the mean absolute percentage error (MAPE) which provides a better idea about the scale of the errors. For the test set $\mathcal{T}_2$, the prediction RMSE and the prediction MAPE are defined as

\begin{align}
    \mathrm{Prediction \; RMSE} &= \sqrt{\frac{1}{\abs{\mathcal{T}_2}}\sum_{x\in\mathcal{T}_2}(\hat\mu(x)-\mu(x))^2}, \\
    \mathrm{Prediction \; MAPE} &= \frac{1}{\abs{\mathcal{T}_2}}\sum_{x\in\mathcal{T}_2} \abs{\frac{\hat\mu(x)-\mu(x)}{\mu(x)}}.
\end{align}

All calculations in this paper are done in RStudio version 1.2.5033 (coupled with R version 3.6.2). Computation of the standard quantile regression estimator $\widehat {\bm\beta}_{\tau;kb}$ of \cref{eq:estimate-kb} is done through the quantreg package by \cite{quantreg}.

\section{Simulation study}
\label{sec:simulation}

The objective of the simulation study in this paper is two-fold. On one hand, we want to observe how well the three different approaches capture the underlying structure of a data generating process (DGP) while on the other, we want to compare the predictive accuracy of the three methods.

To that end, we consider different DGPs (see \Cref{tab:DGP}). Here, $\mu(\cdot)$ and $\sigma(\cdot)$ refer to the mean and variance functions as defined in \cref{eq:sqrm}. First, we consider a SQRM-type structure, where $\x_t$, for $1\leqslant t\leqslant n$, are assumed to be iid, $\mu(\x_t)$ is a linear function and $\sigma^2(\x_t)$ is constant. Second, we keep the same assumptions for $\x_t$ and $\sigma^2(\x_t)$, but consider a nonlinear function for $\mu(\x_t)$. We call it a nonlinear quantile regression model and abbreviate as NLQRM. Next, we do not consider $\sigma^2(\x_t)$ to be constant, but make it proportional to $\abs{\x_t}$, so as to impose potential heteroskedasticity in the model. This model is hereafter termed nonlinear heteroskedastic quantile regression model (NLHQRM). Finally, we relax the iid assumption on $\x_t$ as well, and generate it from an autoregressive process of order 1. This is a general quantile regression model (GQRM). 

\begin{table}[!htb]
    \centering
    \caption{List of DGPs for the simulation study.}
    \label{tab:DGP}
    \begin{tabular}{|lccc|}
    \hline
    DGP & $\x_t$ & $\mu(\x_t)$ & $\sigma^2(\x_t)$ \\
    \hline
    SQRM & iid & linear in $\x_t$ & constant \\
    NLQRM & iid & nonlinear in $\x_t$ & constant \\
    NLHQRM & iid & nonlinear in $\x_t$ & $\propto \abs{\x_t}$ \\
    GQRM & AR(1) & nonlinear in $\x_t$ & $\propto \abs{\x_t}$ \\
    \hline
    \end{tabular}
\end{table}

For all of the above DGPs, we use scalar covariates $\x_t$ for all $1\leqslant t \leqslant n$. For notational clarity, let us use $X_t$ to denote the scalar covariate. For the linear functional form $\mu(X_t)=\beta_0 + \beta_1X_t$, we consider $\beta_0=1$, $\beta_1=0.4$. The constant for $\sigma^2(\cdot)$ is always taken as 1. In the last three DGPs, we take a quadratic form $\mu(X_t) = \beta_0 + \beta_1X_t+  \beta_2X_t^2$ for the nonlinear function in covariate, and use $\beta_0=2.2$, $\beta_1 = 1.7$, $\beta_2=-0.5$ to simulate the data. In case of GQRM, the AR(1) process is generated using an autoregressive coefficient of 0.6, and with innovations coming from iid standard normal distributions.

Similar to the real life applications discussed in \Cref{sec:application}, the index set $\Gamma_t$ for the replicated observations at time $t$ are assumed to be unobserved, but $\abs{\Gamma_t}$ are known. In all of the four cases, we choose a balanced design with $\abs{\Gamma_t} = k$. We observe the results for the three competing methods and compare them for different higher quantiles at probability levels $\tau \in \{0.7,0.8,0.9,0.95\}$. Both $n$ (number of time-points) and $k$ (number of replicates) are varied to check for robustness and to understand the difference between small sample and large sample performances of the three estimators. Most of the choices of the parameter values of this section are in line with the related simulation study conducted by \cite{jana2019improving}.

As mentioned before, we want to find out how well our proposed approach captures the underlying structure of the DGP and also to evaluate the prediction accuracy of the method. Accordingly, for every experiment, we use the first 80\% of the simulated data as the training set and keep last 20\% of the observations as the out-of-sample test set. Then, we fit the models on the training set, and use the results to make predictions for the test set. For every such experiment, the RMSE for the training set and the prediction RMSE are calculated. In order to check for robustness, we repeat every experiment 1000 times and report the mean RMSE and the mean prediction RMSE below. We shall also point out the corresponding prediction MAPE to have a better understanding of the scale of the errors.

Note that for the brevity of the paper, we present a comparative discussion on the performance of the three methods across different DGPs and include detailed results only for SQRM and GQRM. All other simulation results can be found in \Cref{sec:additional-results}. In all these results, KB corresponds to the method that uses \cref{eq:estimate-kb}, JS is the approach that uses \cref{eq:estimate-js}, and NP stands for our proposed approach corresponding to \cref{eq:mu-estimate}.

First, in \Cref{tab:MSE-SQRM} and \Cref{tab:MSE-GQRM}, the mean RMSE of the fitted quantiles for all the simulations for SQRM and GQRM respectively, corresponding to different values of $n$ and $k$, are presented. A couple of interesting phenomena can be observed in these tables. For SQRM, the values are close to each other for all the three methods. For GQRM though, it is evident that the nonparametric method outperforms the other two methods. The RMSE is around 9 times higher in case of KB or JS, and that is true for all values of $n$. Moreover, the results are similar for all quantiles. On the other hand, for our method, we observe that the RMSE decreases substantially as $k$ increases for every choice of $n$ while for same $k$, the errors are decreasing at a slower rate for different values of $n$. This points to the fact that the fit gets much better with more number of replicates. Interestingly, KB and JS method display a similar phenomena for SQRM, but not for GQRM. In fact, even for NLQRM and NLHQRM (refer to \Cref{sec:additional-results}), KB and JS perform similarly as in the case of GQRM. Thus, it is clear that whenever the signal function is assumed to be nonlinear, the performance of the two standard methods suffer and these patterns do not change much even for larger values of $n$ and $k$.

\begin{table}[!hbt]
\centering
\caption{Empirical RMSE (mean taken over 1000 simulations) of fitted quantiles using the three different approaches for different values of $\tau$, $K$ and $n$, when the data are generated from a SQRM.}
\label{tab:MSE-SQRM}
\begin{tabular}{|cc|ccc|ccc|ccc|}
    \hline
    & & \multicolumn{3}{c|}{$n$=30} & \multicolumn{3}{c|}{$n$=50} & \multicolumn{3}{c|}{$n$=100} \\
	$\tau$ & Method & $k$=50 & $k$=100 & $k$=500 & $k$=50 & $k$=100 & $k$=500 & $k$=50 & $k$=100 & $k$=500 \\
	\hline
	0.7 & KB & 0.298 & 0.272 & 0.251 & 0.300 & 0.273 & 0.250 & 0.301 & 0.273 & 0.253 \\ 
        & JS & 0.297 & 0.271 & 0.250 & 0.299 & 0.272 & 0.249 & 0.300 & 0.272 & 0.252 \\ 
        & NP & 0.177 & 0.129 & 0.082 & 0.180 & 0.132 & 0.083 & 0.183 & 0.133 & 0.083 \\ 
    \hline
    0.8 & KB & 0.307 & 0.278 & 0.251 & 0.309 & 0.277 & 0.254 & 0.310 & 0.279 & 0.254 \\ 
        & JS & 0.304 & 0.276 & 0.250 & 0.306 & 0.275 & 0.253 & 0.307 & 0.277 & 0.253 \\ 
        & NP & 0.188 & 0.138 & 0.087 & 0.194 & 0.141 & 0.088 & 0.197 & 0.144 & 0.089 \\ 
    \hline
    0.9 & KB & 0.332 & 0.290 & 0.259 & 0.335 & 0.294 & 0.263 & 0.335 & 0.295 & 0.263 \\ 
        & JS & 0.324 & 0.286 & 0.256 & 0.326 & 0.288 & 0.259 & 0.326 & 0.289 & 0.259 \\ 
        & NP & 0.221 & 0.161 & 0.099 & 0.226 & 0.165 & 0.102 & 0.230 & 0.168 & 0.103 \\ 
    \hline
    0.95 & KB & 0.366 & 0.315 & 0.271 & 0.370 & 0.318 & 0.271 & 0.369 & 0.319 & 0.273 \\ 
        & JS & 0.351 & 0.305 & 0.265 & 0.352 & 0.306 & 0.265 & 0.351 & 0.308 & 0.267 \\ 
        & NP & 0.254 & 0.194 & 0.118 & 0.261 & 0.197 & 0.122 & 0.265 & 0.201 & 0.124 \\ 
	\hline
\end{tabular}
\end{table}

\begin{table}[!hbt]
\centering
\caption{Empirical RMSE (mean taken over 1000 simulations) of fitted quantiles using the three different approaches for different values of $\tau$, $K$ and $n$, when the data are generated from a GQRM.}
\label{tab:MSE-GQRM}
\begin{tabular}{|cc|ccc|ccc|ccc|}
    \hline
    & & \multicolumn{3}{c|}{$n$=30} & \multicolumn{3}{c|}{$n$=50} & \multicolumn{3}{c|}{$n$=100} \\
	$\tau$ & Method & $k$=50 & $k$=100 & $k$=500 & $k$=50 & $k$=100 & $k$=500 & $k$=50 & $k$=100 & $k$=500 \\
	\hline
	0.7 & KB & 1.790 & 1.776 & 1.776 & 1.787 & 1.784 & 1.794 & 1.793 & 1.790 & 1.793 \\ 
        & JS & 1.814 & 1.796 & 1.793 & 1.802 & 1.795 & 1.803 & 1.799 & 1.795 & 1.799 \\ 
        & NP & 0.279 & 0.272 & 0.244 & 0.265 & 0.237 & 0.222 & 0.243 & 0.208 & 0.188 \\ 
    \hline
    0.8 & KB & 1.809 & 1.800 & 1.778 & 1.798 & 1.794 & 1.800 & 1.797 & 1.800 & 1.797 \\ 
        & JS & 1.817 & 1.797 & 1.784 & 1.806 & 1.796 & 1.799 & 1.800 & 1.801 & 1.798 \\ 
        & NP & 0.289 & 0.266 & 0.249 & 0.267 & 0.244 & 0.223 & 0.251 & 0.219 & 0.191 \\ 
    \hline
    0.9 & KB & 1.813 & 1.823 & 1.821 & 1.804 & 1.801 & 1.805 & 1.784 & 1.799 & 1.801 \\ 
        & JS & 1.834 & 1.808 & 1.794 & 1.816 & 1.802 & 1.792 & 1.789 & 1.799 & 1.796 \\ 
        & NP & 0.303 & 0.278 & 0.249 & 0.293 & 0.257 & 0.226 & 0.277 & 0.233 & 0.197 \\ 
    \hline
    0.95 & KB & 1.821 & 1.819 & 1.843 & 1.807 & 1.810 & 1.792 & 1.784 & 1.794 & 1.807 \\ 
        & JS & 1.881 & 1.808 & 1.808 & 1.859 & 1.816 & 1.788 & 1.813 & 1.797 & 1.799 \\ 
        & NP & 0.324 & 0.289 & 0.262 & 0.317 & 0.279 & 0.235 & 0.304 & 0.255 & 0.207 \\
	\hline
\end{tabular}
\end{table}

Turn attention to the prediction performances next. Similar to before, prediction RMSE in different cases for SQRM are depicted in \Cref{tab:predMSE-SQRM} and the same for GQRM are shown in \Cref{tab:predMSE-GQRM}. Here also, we can see that the errors are comparable for the simple model. The prediction RMSE for all three methods are around 0.3, which is equivalent to approximately 10-15\% prediction MAPE. However, for GQRM, the nonparametric approach performs at least three times better than the competitors. NP maintains around 10\% MAPE in this scenario, but the MAPE for KB and JS are more than 30\%. Also, for the NP method, the prediction error diminishes gradually for both $n$ and $k$ whereas it remains stable for the two benchmark methods.

\begin{table}[!hbt]
\centering
\caption{Prediction RMSE (mean taken over 1000 simulations) for the three different approaches for different values of $\tau$, $K$ and $n$, when the data are generated from a SQRM.}
\label{tab:predMSE-SQRM}
\begin{tabular}{|cc|ccc|ccc|ccc|}
    \hline
    & & \multicolumn{3}{c|}{$n$=30} & \multicolumn{3}{c|}{$n$=50} & \multicolumn{3}{c|}{$n$=100} \\
	$\tau$ & Method & $k$=50 & $k$=100 & $k$=500 & $k$=50 & $k$=100 & $k$=500 & $k$=50 & $k$=100 & $k$=500 \\
	\hline
	0.7 & KB & 0.311 & 0.283 & 0.259 & 0.303 & 0.279 & 0.258 & 0.305 & 0.277 & 0.255 \\ 
        & JS & 0.310 & 0.283 & 0.258 & 0.302 & 0.278 & 0.257 & 0.304 & 0.276 & 0.255 \\ 
        & NP & 0.201 & 0.148 & 0.096 & 0.192 & 0.144 & 0.093 & 0.190 & 0.139 & 0.088 \\ 
    \hline
    0.8 & KB & 0.322 & 0.288 & 0.259 & 0.318 & 0.285 & 0.262 & 0.313 & 0.282 & 0.257 \\ 
        & JS & 0.319 & 0.286 & 0.259 & 0.315 & 0.283 & 0.260 & 0.310 & 0.280 & 0.255 \\ 
        & NP & 0.214 & 0.157 & 0.102 & 0.208 & 0.154 & 0.099 & 0.205 & 0.149 & 0.093 \\ 
    \hline
    0.9 & KB & 0.349 & 0.306 & 0.267 & 0.341 & 0.299 & 0.266 & 0.339 & 0.297 & 0.264 \\ 
        & JS & 0.339 & 0.300 & 0.263 & 0.331 & 0.293 & 0.262 & 0.330 & 0.292 & 0.260 \\ 
        & NP & 0.250 & 0.185 & 0.116 & 0.244 & 0.179 & 0.112 & 0.239 & 0.175 & 0.108 \\ 
    \hline
    0.95 & KB & 0.380 & 0.332 & 0.281 & 0.384 & 0.324 & 0.273 & 0.374 & 0.324 & 0.276 \\ 
        & JS & 0.364 & 0.323 & 0.276 & 0.360 & 0.312 & 0.266 & 0.358 & 0.311 & 0.269 \\ 
        & NP & 0.288 & 0.224 & 0.137 & 0.284 & 0.217 & 0.132 & 0.276 & 0.211 & 0.130 \\ 
	\hline
\end{tabular}
\end{table}

\begin{table}[!hbt]
\centering
\caption{Prediction RMSE (mean taken over 1000 simulations) for the three different approaches for different values of $\tau$, $K$ and $n$, when the data are generated from a GQRM.}
\label{tab:predMSE-GQRM}
\begin{tabular}{|cc|ccc|ccc|ccc|}
    \hline
    & & \multicolumn{3}{c|}{$n$=30} & \multicolumn{3}{c|}{$n$=50} & \multicolumn{3}{c|}{$n$=100} \\
	$\tau$ & Method & $k$=50 & $k$=100 & $k$=500 & $k$=50 & $k$=100 & $k$=500 & $k$=50 & $k$=100 & $k$=500 \\
	\hline
	0.7 & KB & 1.879 & 1.889 & 1.878 & 1.860 & 1.839 & 1.840 & 1.824 & 1.822 & 1.832 \\ 
        & JS & 1.911 & 1.906 & 1.898 & 1.877 & 1.850 & 1.853 & 1.829 & 1.829 & 1.836 \\ 
        & NP & 0.550 & 0.498 & 0.461 & 0.458 & 0.418 & 0.409 & 0.360 & 0.342 & 0.329 \\ 
    \hline
    0.8 & KB & 1.900 & 1.894 & 1.902 & 1.860 & 1.835 & 1.850 & 1.816 & 1.823 & 1.829 \\ 
        & JS & 1.933 & 1.898 & 1.911 & 1.871 & 1.847 & 1.861 & 1.826 & 1.826 & 1.832 \\ 
        & NP & 0.508 & 0.530 & 0.500 & 0.448 & 0.420 & 0.411 & 0.374 & 0.336 & 0.315 \\ 
    \hline
    0.9 & KB & 1.936 & 1.896 & 1.929 & 1.847 & 1.867 & 1.850 & 1.814 & 1.815 & 1.823 \\ 
        & JS & 1.941 & 1.906 & 1.904 & 1.894 & 1.880 & 1.853 & 1.827 & 1.823 & 1.829 \\ 
        & NP & 0.593 & 0.526 & 0.527 & 0.488 & 0.453 & 0.414 & 0.380 & 0.365 & 0.310 \\ 
    \hline
    0.95 & KB & 1.913 & 1.921 & 1.900 & 1.848 & 1.825 & 1.866 & 1.811 & 1.805 & 1.820 \\ 
        & JS & 1.981 & 1.953 & 1.901 & 1.929 & 1.856 & 1.882 & 1.856 & 1.821 & 1.822 \\ 
        & NP & 0.604 & 0.543 & 0.525 & 0.505 & 0.491 & 0.433 & 0.413 & 0.377 & 0.344 \\ 
	\hline
\end{tabular}
\end{table}

A brief discussion of the results for the other two DGPs (NLQRM and NLHQRM) is of the essence to end this section. For both of these processes, NP method shows an RMSE of around one-fourth of what we see in case of KB or JS methods, and that is true for all combinations of $n$, $k$ and $\tau$. There also, the RMSE of the nonparametric method improves substantially with larger values of $k$ and the performance is good for all quantiles. Prediction-wise as well, NP beats KB and JS by and large, even though the latter two record better accuracy than what we see in GQRM. All in all, it can be said that the proposed nonparametric approach performs consistently better than the existing approaches whenever the DGP deviates from a simple quantile regression model. In fact, we find that for both short term and long term forecasting, the proposed approach has a decided advantage over the standard methods. We omit these results for the interest of space. Refer to \Cref{sec:additional-results} for detailed results for all of the DGPs.

\section{Application}\label{sec:application}

\subsection{Tropical cyclone windspeed data} 
\label{subsec:cyclone}

As the first real application of the proposed method, we re-analyse the tropical cyclone data considered in \cite{elsner2008increasing}. This data has been made publicly available by \cite{jagger2006climatology} and can be downloaded from the website \href{https://myweb.fsu.edu/jelsner/temp/Data.html}{https://myweb.fsu.edu/jelsner/temp/Data.html}. This satellite based data consists of lifetime maximum wind speed (hereafter denoted as Wmax, evaluated in meters per second) for all tropical cyclones in the National Hurricane Center (NHC) best track, recorded over the period of 1899 to 2009. As covariates in the model, we consider annually averaged global Sea-Surface Temperature (SST), and Southern Oscillation Index (SOI). In addition, an intercept term and a linear trend are also used as covariates. It is worth mention that both SST and SOI have been extensively used in modeling cyclones, cf.\ \cite{nicholls1998recent} and \cite{kang2019contribution}. In this data, the two covariates are observed yearly while for every year there are multiple observations for Wmax and thus, it is under the framework of our study. Number of replications per year vary up to 47. However, recall that our theory is developed under the assumption of sufficient replicates for every time point. Keeping that in mind, we remove the years where the number of replicates is less than 10. That leads to a total of 104 years of data. Finally, as our main focus is on higher wind speed storms (as in \cite{elsner2008increasing}) which cause major damages, we consider quantiles at the upper probability levels  $\tau \in \{0.7, 0.8, 0.9, 0.95\}$.

In the main analysis, Wmax values are transformed to the square-root scale. Several authors (e.g.\ \cite{brown1984time}, \cite{torres2005forecast}) have argued in detail why this transformation is appropriate in this context. Next, in order to assess both the aspects of inference and prediction, we split the data into training and test sets. Similar to the simulation studies, approximately 20\% of the total, that is the last 21 years of data, are kept as the test set. All three methods are fit to the training data using the covariates and then predictions are made for the test set. Accuracy metrics are then computed and presented in \Cref{tab:cyclone-MSE}. 

\begin{table}[!hbt]
\centering
\caption{RMSE for the fitted quantiles and the overall prediction accuracy (RMSE and MAPE) for the test set, corresponding to different values of $\tau$, from the cyclone data.} 
\label{tab:cyclone-MSE}
\begin{tabular}{|c|ccc|ccc|ccc|}
    \hline
    & \multicolumn{3}{c|}{Training RMSE} & \multicolumn{3}{c|}{Prediction RMSE} & \multicolumn{3}{c|}{Prediction MAPE} \\
    $\tau$ & KB & JS & NP & KB & JS & NP & KB & JS & NP \\
    \hline
    0.7  & 0.889 & 0.903 & 0.801 & 0.905 & 0.983 & 0.899 & 8.66\% & 9.40\% & 8.44\% \\  
    0.8  & 1.006 & 1.029 & 0.888 & 1.218 & 1.241 & 0.931 & 11.59\% & 11.56\% & 8.60\% \\ 
    0.9  & 1.020 & 0.976 & 0.864 & 1.222 & 1.153 & 0.854 & 10.36\% & 9.72\% & 7.16\% \\ 
    0.95 & 1.097 & 1.445 & 0.807 & 1.185 & 3.055 & 0.770 & 9.87\% & 27.47\% & 5.65\% \\ 
    \hline
\end{tabular}
\end{table}

We see that the nonparametric approach records uniformly better RMSE in both the training set and the test set. For $\tau=0.7$, the differences between NP and the other two methods are about 0.1, but the same increases substantially for higher quantiles. For instance, in case of 95\% quantile, the prediction RMSE for NP is about 35\% better than that of KB method, whereas the JS method fails miserably. The prediction MAPE values are within 10\% in all cases, thereby suggesting that the scales of the errors are not too high. It is necessary to point out that the MAPE for the NP approach decreases for higher quantiles. At $\tau=0.95$, it is only 5.65\%.

\begin{figure}[!hbt]
\centering
\caption{Modelled series for the transformed wind-speed data for different quantiles in the three different approaches. The dotted line represents the beginning of the test period.}
\label{fig:cyclone-fit}
\includegraphics[width=0.9\textwidth,keepaspectratio]{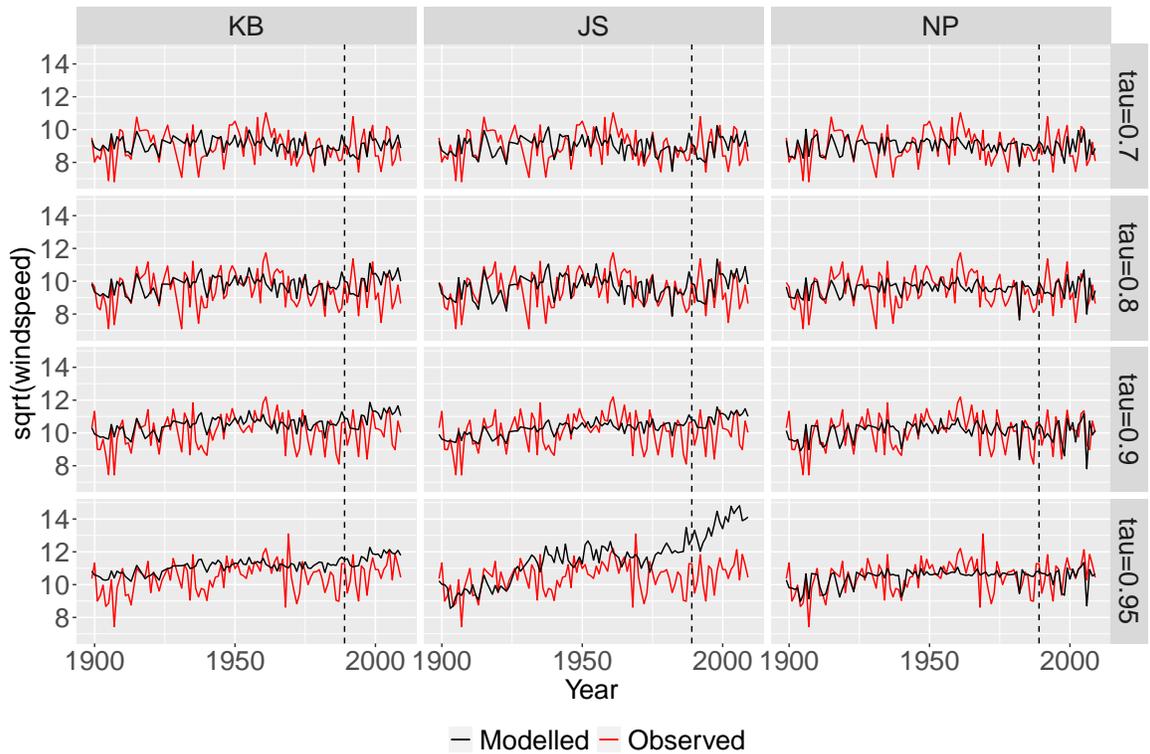}
\end{figure}

Digging deeper into the results, we find that both KB and JS have a tendency to overestimate the mean function. They also fail to capture the peaks and troughs of the data properly, which is not true for our proposed method. It can be observed from \Cref{fig:cyclone-fit} where the true series and the fitted series for the three methods are presented. In the figure, results corresponding to the three methods are presented column-wise while the graphs for different quantiles are presented row-wise. The dotted line in each graph represents the beginning of the test period. For both $\tau=0.7$ and 0.8, all three methods tend to perform at par with each other, but for the extreme quantiles, only the nonparametric approach makes great predictions. It firmly establishes the superiority of the proposed method for higher values of $\tau$.

\begin{figure}[!hbt]
\centering
\caption{Contour plot of the estimated mean function (for the wind-speed data) in the proposed nonparametric method for different choices of the covariate values, corresponding to different quantiles.}
\label{fig:cyclone-npeffect}
\includegraphics[width=\textwidth,keepaspectratio]{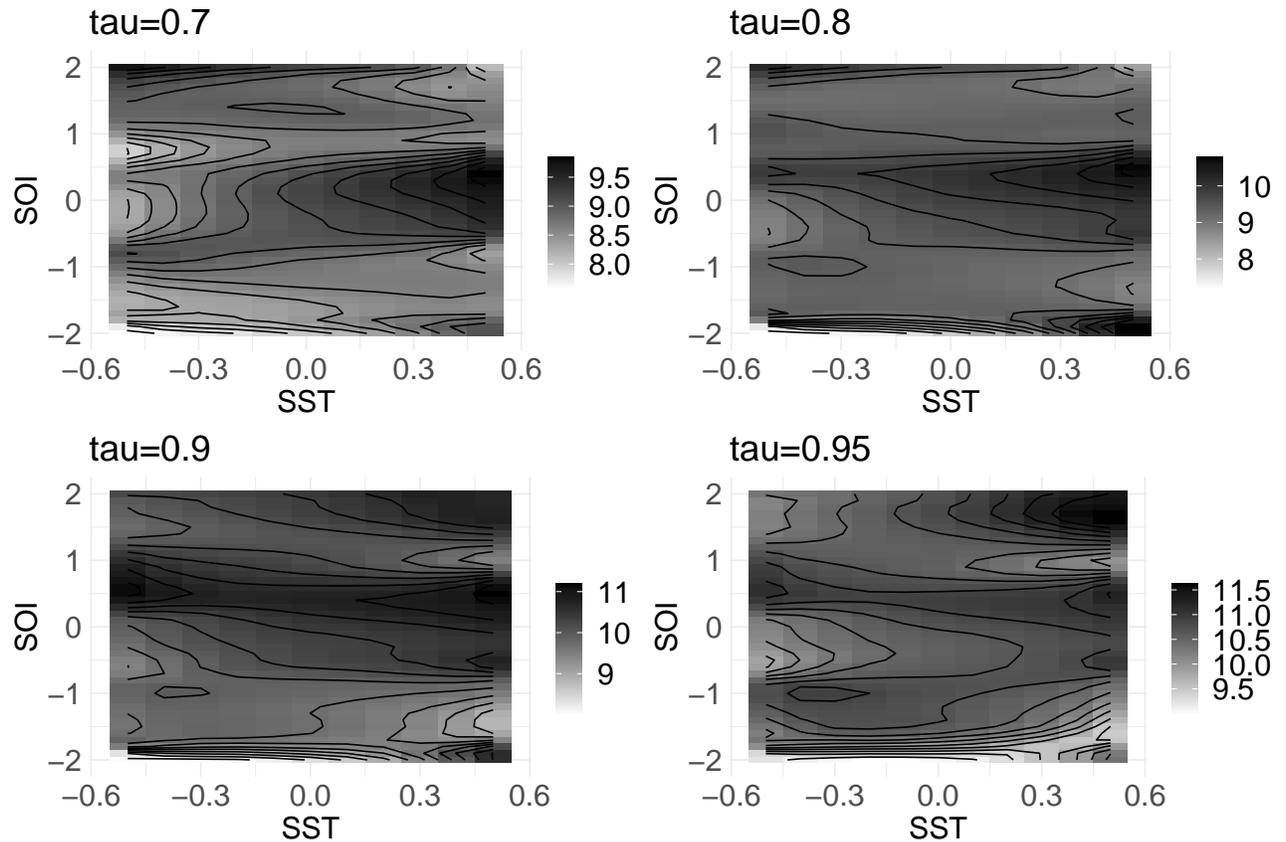}
\end{figure}

To further investigate the results of the nonparametric approach, we take a look at the contour plot (\Cref{fig:cyclone-npeffect}) of the estimated $\mu(x)$ values for different combinations of SST and SOI values at a certain time-point. Remember that both KB and JS assume linearity in the effect of the covariates. However, the contour plot clearly provides strong support in favor of the hypothesis that the mean function is nonlinear in both the regressors at all quantile levels. Along a similar line, we find out that the nonparametric method identifies a nonlinear trend in the data. Moreover, $\sigma(\cdot)$ values for different choices of the covariate levels are estimated to be widely varying. At all quantile levels, it ranges between 0 and 2, which substantiates the use of a general functional form for the conditional variance in our model. Overall, the nonlinear effects of all covariates and the heteroskedastic nature of the variance function are the most plausible explanation of why the proposed approach outperforms the benchmark methods in this application.

Our next analysis is focused to check for robustness of the proposed method in terms of forecasting performance. To that end, we consider different 10-year long forecast periods and evaluate the prediction accuracy metrics in those cases. It should be pointed out that we examined the forecast accuracy for shorter windows as well, and the conclusions were identical. In \Cref{tab:cyclone-MSE-diffperiod}, results for five different forecast periods, namely 1991-2000 to 1999-2008 (consecutive periods have two-year shifts), are presented. We notice that for the first three periods, KB method provides best prediction for $\tau=0.7$, albeit the other two methods are not too worse. For the last two periods in this case, with more data available in the training set, the nonparametric predictions tend to be more accurate. In case of the other quantiles, NP approach consistently performs better. Akin to the previously discussed result, we see that for the largest quantile, the proposed method is by far the best among the competing ones. The MAPE values in case of $\tau=0.95$ are only about 4 to 6\% in NP, whereas KB and JS usually record close to or more than 10\% MAPE. It can be inferred that the benchmark methods work well for lower quantiles, but for the relatively extremes, our proposal has a decided advantage.

\begin{table}[!hbt]
\centering
\caption{Prediction RMSE (MAPE given in parentheses) for different forecast periods (10-year window) in the cyclone data.}
\label{tab:cyclone-MSE-diffperiod}
\begin{tabular}{|cc|ccccc|}
  \hline
  $\tau$ &  & 1991-2000 & 1993-2002 & 1995-2004 & 1997-2006 & 1999-2008 \\ 
  \hline
  0.7 & KB & 0.99 (8.4\%) & 0.815 (8.4\%) & 0.855 (8.7\%) & 0.961 (10.7\%) & 0.976 (11.2\%) \\ 
  & JS & 1.069 (9.1\%) & 0.865 (8.9\%) & 0.885 (9.1\%) & 1.018 (11.4\%) & 0.982 (11.3\%) \\ 
  & NP & 0.991 (8.9\%) & 0.959 (9.8\%) & 0.947 (9.4\%) & 1.002 (10.5\%) & 0.941 (9.9\%) \\ 
  \hline
  0.80 & KB & 1.13 (10.2\%) & 1.346 (13.6\%) & 1.257 (11.9\%) & 1.369 (14\%) & 1.373 (14.4\%) \\ 
   & JS & 1.23 (10.2\%) & 1.229 (11.9\%) & 1.201 (11.4\%) & 1.366 (14.1\%) & 1.309 (13.9\%) \\ 
   & NP & 1.057 (8.8\%) & 1.013 (9.6\%) & 0.807 (7.2\%) & 0.924 (8.8\%) & 0.963 (9.5\%) \\ 
  \hline
  0.90 & KB & 1.183 (10.1\%) & 1.267 (11.4\%) & 1.128 (9.3\%) & 1.239 (10.3\%) & 1.151 (9.9\%) \\ 
    & JS & 1.135 (9.8\%) & 1.19 (10.7\%) & 1.062 (8.8\%) & 1.179 (9.8\%) & 1.142 (9.7\%) \\ 
    & NP & 0.914 (7.5\%) & 0.831 (7\%) & 0.69 (5.6\%) & 0.902 (7.5\%) & 0.858 (6.5\%) \\ 
  \hline
  0.95 & KB & 1.193 (10.3\%) & 1.311 (11.4\%) & 1.031 (8.4\%) & 0.819 (6.5\%) & 0.839 (7\%) \\ 
    & JS & 2.9 (26\%) & 3.23 (30.1\%) & 3.164 (28.6\%) & 3.195 (28.3\%) & 3.246 (29.1\%) \\ 
    & NP & 0.826 (6.6\%) & 0.693 (5.4\%) & 0.641 (4.4\%) & 0.763 (5.8\%) & 0.569 (4.3\%) \\ 
  \hline
\end{tabular}
\end{table}

\subsection{Air pollution data from Bengaluru, India}
\label{subsec:pollution}

In this section, we analyze air pollution data from Bengaluru, the capital of the state of Karnataka in India. Bengaluru is one of the fastest growing city in the country. In fact, according to a recent report (\cite{dealroom}), it is the fastest growing mature tech ecosystem in the world since 2016. With this industrial growth and a subsequent boom in population, there has been considerable challenge to maintain good air quality. Naturally, it is of paramount interest to study and forecast the pollution levels in the city, see \cite{abhilash2018time} and \cite{guttikunda2019air} for example.  

The data we analyze in this paper is publicly available from the Central Pollution Control Board (website: \href{https://cpcb.nic.in/}{https://cpcb.nic.in/}), an official portal of the Government of India. Our main variable in this discussion is related to the particulate matters (PM) with an aerodynamic diameter of less than 2.5 microns. It is measured in micrograms per cubic meter of air. Hereafter, we denote this variable as $\mathrm{PM2.5}$. Several epidemiological studies, such as \cite{jerrett2005spatial} and \cite{thurston2016ischemic}, have established that $\mathrm{PM2.5}$ is linked to a range of serious cardiovascular, respiratory, and other health problems. It is in fact one of the main variables in the study of air pollution. 

In case of the data considered in the paper, the measures of $\mathrm{PM2.5}$ are obtained at different hours of a single day from different stations of Bengaluru. As the information of both aspects (hour and location) are not available, it falls within the framework of the current study, where the observations obtained on a single day are considered to be replicated observations at each time point, i.e.\ on a single day. We use the data from 20th March 2015 to 1st July 2020. For the main analysis, first we remove the cases where the recorded values of the $\mathrm{PM2.5}$ observations are negative, possibly due to erroneous data preparation. We also remove the days where the numbers of replicates are less than a cutoff. That leaves us a total of 1622 days of data. Next, in line with the previous analyses, we use the first 80\% time-points in the training set and keep last 20\% as the test set. Consequently, we have approximately one year of data in the test set and it helps us to judge the long-term prediction accuracy in the context of the pollution problem.

While carrying out the main analysis, the $\mathrm{PM2.5}$ observations are converted to logarithmic scale using the transformation $x\to \log(1+x)$. This is a common pre-processing step in the air pollution literature, cf.\ \cite{smith2003spatiotemporal}. Moreover, it is imperative to point out that this problem is different from \Cref{subsec:cyclone} as we have only time covariates in the data. In order to capture the time-dependence appropriately, we use monthly seasonalities and a common intercept term. We also explored the possibility of including linear and quadratic trends and weekly seasonality terms, but they appeared not to improve the results and are hence omitted from the model. 

\begin{figure}[!hbt]
\centering
\caption{Modelled series for the $\log(\mathrm{PM}2.5)$ data for different quantiles in the three different approaches.}
\label{fig:pollution-fit}
\includegraphics[width=0.8\textwidth,keepaspectratio]{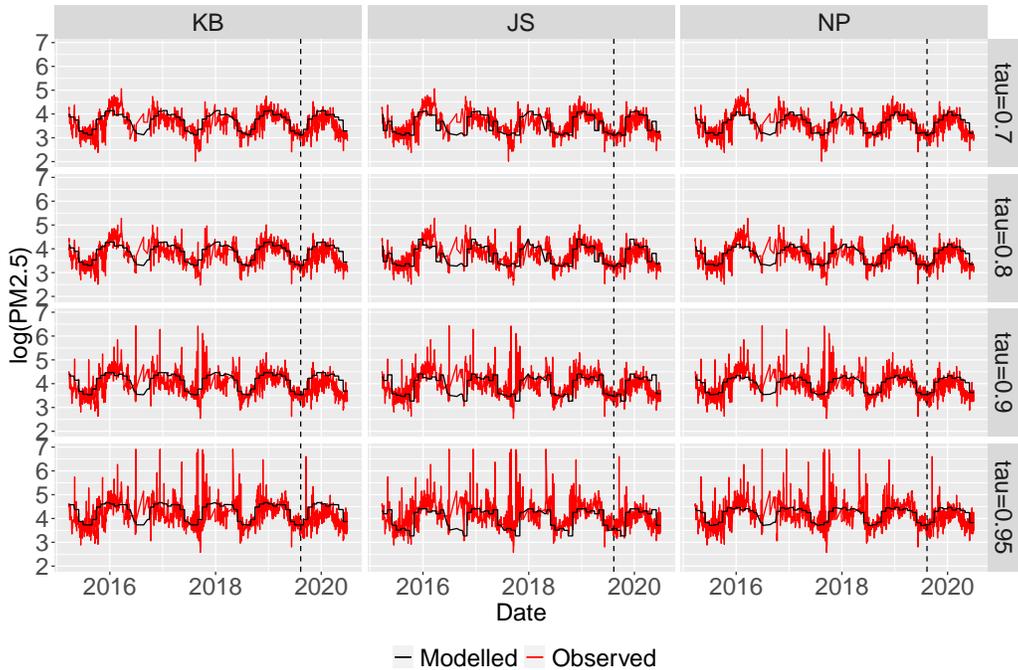}
\end{figure}

As before, four different quantiles ($\tau = 0.7, 0.8, 0.9, 0.95$) are considered. For every quantile, all three methods are applied to the training data and predictions are made for the entire test set. We calculate the RMSE from the training set and the prediction RMSE and MAPE from the test set. These values are presented in \Cref{tab:pollution-MSE}. We also plot the entire series along with the modelled series for different methods in \Cref{fig:pollution-fit}. 

\begin{table}[!hbt]
\centering
\caption{RMSE for the fitted quantiles and the overall prediction accuracy (RMSE and MAPE) for the test set, corresponding to different values of $\tau$, from the air pollution data.}
\label{tab:pollution-MSE}
\begin{tabular}{|c|ccc|ccc|ccc|}
    \hline
    & \multicolumn{3}{c|}{Training RMSE} & \multicolumn{3}{c|}{Prediction RMSE} & \multicolumn{3}{c|}{Prediction MAPE} \\
    $\tau$ & KB & JS & NP & KB & JS & NP & KB & JS & NP \\
    \hline
    0.7  & 0.345 & 0.392 & 0.328 & 0.446 & 0.375 & 0.376 & 11.19\% & 8.97\% & 9.09\% \\ 
    0.8  & 0.350 & 0.401 & 0.332 & 0.464 & 0.394 & 0.380 & 11.34\% & 8.97\% & 8.95\% \\ 
    0.9  & 0.421 & 0.481 & 0.401 & 0.510 & 0.478 & 0.410 & 12.00\% & 10.36\% & 9.35\% \\ 
    0.95 & 0.495 & 0.579 & 0.470 & 0.584 & 0.500 & 0.466 & 12.92\% & 9.68\% & 9.79\% \\ 
    \hline
\end{tabular}
\end{table}

From both \Cref{tab:pollution-MSE} and \Cref{fig:pollution-fit}, it is clear that all three methods fit the data well. The RMSE values are not too different for the training set, although the values corresponding to the JS approach are marginally higher. This pattern, however, does not remain the same in the prediction part. For lower quantiles, we can see that both NP and JS record slightly higher prediction accuracy while for higher quantiles, our proposed approach beats the other two methods by and large. We emphasize that these differences in the RMSE values are minute and the three methods can be thought to be performing at par with each other. It is in stark contrast with the previous application. A probable explanation is that without any exogenous regressor, all methods would perform similarly. We also hypothesize that the data is more structured in this case and hence, the nonparametric approach provides only minor improvement than the standard approaches.

\begin{figure}[!hbt]
\centering
\caption{Comparison of the coefficient estimates in the three methods for the pollution data, corresponding to different quantiles.}
\label{fig:pollution-npeffect}
\includegraphics[width=0.8\textwidth,keepaspectratio]{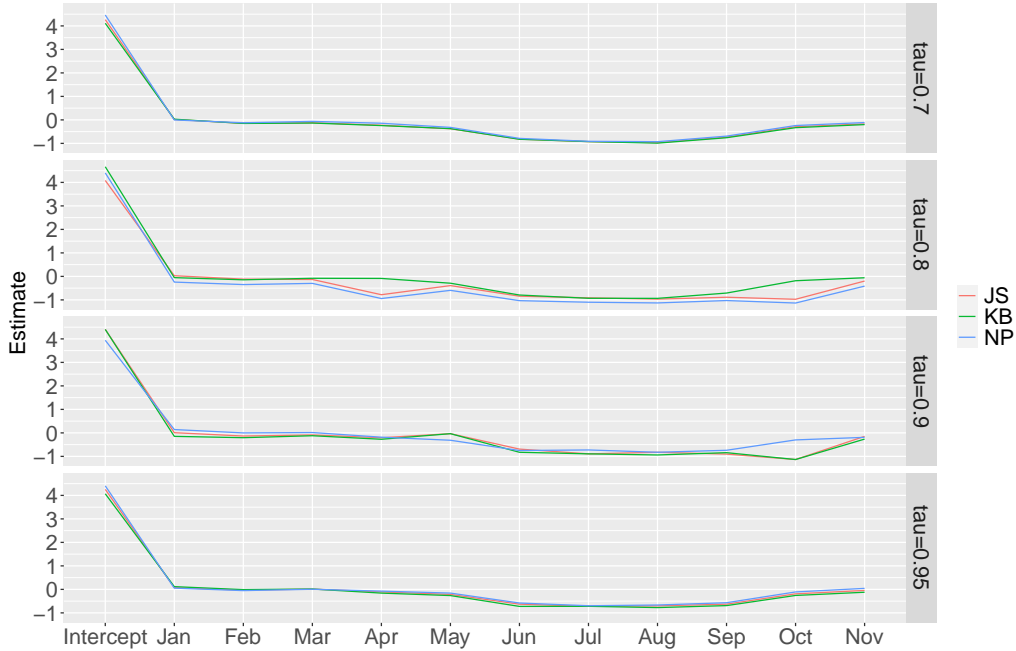}
\end{figure}

To better explain the above, observe that the only covariates we use in this analysis are the intercept term and the monthly indicators, which capture the seasonality pattern of the data. Since all of these regressors are binary and non-random quantities, the nonparametric method is in essence similar to KB or JS, as all of them just focus on estimating the effect sizes of the dummy indicators. We present the estimated coefficients in this regard, in \Cref{fig:pollution-npeffect}. From the graph, is evident that all three methods return almost identical estimates for all of the covariates. This automatically justifies why the methods perform similarly in this application.

As a last piece of the analysis, in the same spirit as before, we consider different forecast periods to check for robustness. Here, keeping the problem in mind, we explore the performance of the methods for both short-term and long-term prediction. In the context of pollution forecasting, both aspects are important for policymakers and climatologists. In case of the long-term prediction (moving windows of one year) problem, akin to the previous application, we find no discernible difference in the performance of the three methods. Thus, identical conclusions (as in \Cref{tab:pollution-MSE} and \Cref{fig:pollution-fit}) can be drawn for different one-year time windows. However, if we consider one-month windows, mixed results are obtained. Consult \Cref{tab:pollution-MSE-diffperiod} in this regard. Prediction RMSE and MAPE for the months of January to June in 2020 are displayed. It is evident that there are a couple of months where NP method turns out to be the best, whereas for others JS is producing the best results. KB method is also showing acceptable accuracy. Furthermore, contrary to the earlier finding, we see that the precision of the three methods in different quantiles do not follow any particular pattern here. Hence, overall, the results reconfirm that without the presence of stochastic covariates, all methods can forecast well both in the short-term and in the long-term.

\begin{table}[!hbt]
\centering
\caption{Prediction RMSE (MAPE given in parentheses) for different forecast periods (one-month window) in the air pollution data.}
\label{tab:pollution-MSE-diffperiod}
\small
\begin{tabular}{|cc|cccccc|}
  \hline
  $\tau$ &  & Jan 2020 & Feb 2020 & Mar 2020 & Apr 2020 & May 2020 & Jun 2020 \\ 
  \hline
  0.7 & KB & 0.4 (9\%) & 0.214 (4.8\%) & 0.383 (9.2\%) & 0.586 (16.5\%) & 0.564 (17.3\%) & 0.342 (9.7\%) \\ 
   & JS & 0.385 (8.6\%) & 0.216 (4.8\%) & 0.356 (8.3\%) & 0.264 (6.8\%) & 0.516 (15.7\%) & 0.304 (8.3\%) \\ 
   & NP & 0.358 (8\%) & 0.199 (4.4\%) & 0.365 (8.7\%) & 0.493 (13.3\%) & 0.458 (13.7\%) & 0.268 (7.1\%) \\ 
  \hline
  0.8 & KB & 0.416 (8.9\%) & 0.245 (5.4\%) & 0.402 (9.5\%) & 0.611 (16.7\%) & 0.579 (17.2\%) & 0.357 (10.2\%) \\ 
   & JS & 0.331 (6.8\%) & 0.204 (4.4\%) & 0.363 (8.2\%) & 0.254 (6.1\%) & 0.503 (14.8\%) & 0.294 (7.9\%) \\ 
   & NP & 0.345 (7.1\%) & 0.205 (4.4\%) & 0.359 (8.2\%) & 0.502 (13.3\%) & 0.5 (14.7\%) & 0.275 (7.3\%) \\ 
  \hline
  0.9 & KB & 0.423 (8.8\%) & 0.295 (6.3\%) & 0.449 (10.2\%) & 0.699 (18.5\%) & 0.655 (18.6\%) & 0.397 (11\%) \\ 
  & JS & 0.287 (5.6\%) & 0.201 (4.1\%) & 0.36 (8\%) & 0.518 (13.2\%) & 0.883 (25.2\%) & 0.321 (8.6\%) \\ 
  & NP & 0.328 (6.4\%) & 0.213 (4.3\%) & 0.351 (7.7\%) & 0.546 (14\%) & 0.557 (15.7\%) & 0.357 (9.8\%) \\ 
  \hline
  0.95 & KB & 0.445 (9.1\%) & 0.319 (6.6\%) & 0.495 (11\%) & 0.772 (19.5\%) & 0.729 (19.7\%) & 0.442 (11.7\%) \\ 
  & JS & 0.304 (5.9\%) & 0.184 (3.8\%) & 0.309 (6.2\%) & 0.43 (10.4\%) & 0.726 (19.7\%) & 0.354 (8.7\%) \\ 
  & NP & 0.33 (6.3\%) & 0.208 (4.2\%) & 0.363 (7.7\%) & 0.546 (13.3\%) & 0.606 (16.3\%) & 0.412 (10.8\%) \\
  \hline
\end{tabular}
\end{table}

\section{Concluding remarks}
\label{sec:conclusion}

In this article, we discuss new ways of modelling time series data with replicated observations, a type of data fairly common in the context of environmental studies. These type of climatological data often exhibit nonlinear mean and heteroskedastic patterns. To model such data, we have proposed a nonparametric approach that leverages sample quantiles of the response variable to estimate quantile function in the presence of covariates. The nonparametric method is kernel based. For a very general framework, under some regularity conditions, we have derived the asymptotic distribution and asymptotic confidence interval of the estimated mean level of the quantile function. We also provided necessary conditions to ensure the consistency of the variance estimates. 

Through a detailed simulation study, we have investigated the small sample performance of the three competing methods. Varieties of data generating processes have been considered, ranging from simple linear homoskedastic model to nonlinear (in covariate) heteroskedastic model with time-dependent covariate process and other models in between. We considered different sample sizes and different set of replications reflecting resemblance to the real datasets analyzed in the paper. The performance of the methods are measured with respect to their ability to capture the signal of the data generating process and to evaluate their predictive performances at different future simulated time points. It is observed that for simple linear and homoskedastic data generating process, the performances of the three methods are comparable, but the nonparametric method demonstrates far better performance in both the modelling and the prediction when the data generating process deviates from simplistic model assumptions. The proposed method also exhibits the tendency to capture the information obtained from replicated data with improving efficiency as the number of replication increases, indicating better exploitation of the information in the replicated data. We emphasize that the parametric methods perform at par with the nonparametric approach when the data are generated under the condition that the response does not depend on time-dependent covariates. Otherwise, there is a discernible difference in the efficacy of the proposed method. Interestingly, this difference in the accuracy escalates further for extreme quantiles. The same is observed in the real data analysis as well. For the first application in \Cref{subsec:cyclone}, where the covariates are time-dependent processes, the gain from nonparametric method is substantial. Contrarily, for the application in \Cref{subsec:pollution}, where we do not have information on any time-dependent covariate, this gain is not observed. Overall, the results support the superiority of the proposed method for higher quantiles in general and it can be used in various climate studies.



Let us conclude the paper with some interesting future directions to this work. Considering that the proposed nonparametric method performs better at reasonably extreme quantiles, it is worth exploring the performance in estimating more extreme situations. Estimation of high return level of random events which cause  climate extremes is an important and pressing question which focuses on estimation of probability of rare events based on limited data. Improvising the proposed non-parametric method through a suitable mixture kernel by bringing in heavy tailed component would be a suitable method to develop for the estimation of very extreme quantiles. Another potential extension of the current work would be the inclusion of high or infinite dimensional covariates, for instance functions or images, to the model. One case in point is the modelling of the distribution of pivotal environmental variables (such as precipitation regions, as was done in \cite{Jana-JASA}) at different quantiles using satellite based multiple images as covariates. One can do it through dimension reduction in the covariate space, in conjunction with similar nonparametric techniques. We plan to investigate such problems in future. 


\section*{Funding sources}


The research of the second author is partially supported by the Alan Turing Institute -- Lloyd's Register Foundation Programme on Data-Centric Engineering.

\bibliographystyle{plainnat}
\bibliography{references}

\begin{thebibliography}{51}
\providecommand{\natexlab}[1]{#1}
\providecommand{\url}[1]{\texttt{#1}}
\expandafter\ifx\csname urlstyle\endcsname\relax
  \providecommand{\doi}[1]{doi: #1}\else
  \providecommand{\doi}{doi: \begingroup \urlstyle{rm}\Url}\fi

\bibitem[Abhilash et~al.(2018)Abhilash, Thakur, Gupta, and
  Sreevidya]{abhilash2018time}
MSK Abhilash, Amrita Thakur, Deepa Gupta, and B~Sreevidya.
\newblock {Time series analysis of air pollution in Bengaluru using ARIMA
  model}.
\newblock In \emph{Ambient Communications and Computer Systems}, pages
  413--426. Springer, 2018.
\newblock \doi{10.1007/978-981-10-7386-1_36}.

\bibitem[Bercu et~al.(2019)Bercu, Capderou, and
  Durrieu]{bercu2019nonparametric}
Bernard Bercu, Sami Capderou, and Gilles Durrieu.
\newblock A nonparametric statistical procedure for the detection of marine
  pollution.
\newblock \emph{Journal of Applied Statistics}, 46\penalty0 (1):\penalty0
  119--140, 2019.

\bibitem[Brown et~al.(1984)Brown, Katz, and Murphy]{brown1984time}
Barbara~G Brown, Richard~W Katz, and Allan~H Murphy.
\newblock Time series models to simulate and forecast wind speed and wind
  power.
\newblock \emph{Journal of Applied Meteorology and Climatology}, 23\penalty0
  (8):\penalty0 1184--1195, 1984.

\bibitem[Cai(2002)]{cai2002regression}
Zongwu Cai.
\newblock Regression quantiles for time series.
\newblock \emph{Econometric theory}, pages 169--192, 2002.

\bibitem[Chen et~al.(2021)Chen, Bertke, and Curwin]{chen2021quantile}
I-Chen Chen, Stephen~J Bertke, and Brian~D Curwin.
\newblock Quantile regression for exposure data with repeated measures in the
  presence of non-detects.
\newblock \emph{Journal of Exposure Science \& Environmental Epidemiology},
  pages 1--10, 2021.
\newblock \doi{10.1038/s41370-021-00345-1}.

\bibitem[Dabo-Niang and Laksaci(2012)]{dabo2012nonparametric}
Sophie Dabo-Niang and Ali Laksaci.
\newblock Nonparametric quantile regression estimation for functional dependent
  data.
\newblock \emph{Communications in statistics-Theory and methods}, 41\penalty0
  (7):\penalty0 1254--1268, 2012.

\bibitem[Ding et~al.(2017)Ding, Schweiger, L’Heureux, Battisti, Po-Chedley,
  Johnson, Blanchard-Wrigglesworth, Harnos, Zhang, Eastman, and
  Steig]{ding2017influence}
Qinghua Ding, Axel Schweiger, Michelle L’Heureux, David~S Battisti, Stephen
  Po-Chedley, Nathaniel~C Johnson, Eduardo Blanchard-Wrigglesworth, Kirstin
  Harnos, Qin Zhang, Ryan Eastman, and Eric~J Steig.
\newblock Influence of high-latitude atmospheric circulation changes on
  summertime arctic sea ice.
\newblock \emph{Nature Climate Change}, 7\penalty0 (4):\penalty0 289--295,
  2017.

\bibitem[Elsner et~al.(2008)Elsner, Kossin, and Jagger]{elsner2008increasing}
James~B Elsner, James~P Kossin, and Thomas~H Jagger.
\newblock The increasing intensity of the strongest tropical cyclones.
\newblock \emph{Nature}, 455\penalty0 (7209):\penalty0 92--95, 2008.

\bibitem[Embrechts et~al.(1997)Embrechts, Klüppelberg, and
  Mikosch]{Embrechts1997}
Paul. Embrechts, Claudia. Klüppelberg, and Thomas Mikosch.
\newblock \emph{Modelling Extremal Events}.
\newblock Berlin: Springer, 1997.

\bibitem[Fan and Gijbels(2018)]{fan2018local}
Jianqing Fan and Irene Gijbels.
\newblock \emph{Local polynomial modelling and its applications: monographs on
  statistics and applied probability 66}.
\newblock Routledge, 2018.

\bibitem[Fernández et~al.(2004)Fernández, Redden, Pietrobelli, and
  Allison]{Fernandez_2004}
José~R. Fernández, David~T. Redden, Angelo Pietrobelli, and David~B. Allison.
\newblock Waist circumference percentiles in nationally representative samples
  of african-american, european-american, and mexican-american children and
  adolescents.
\newblock \emph{The Journal of Pediatrics}, 145\penalty0 (4):\penalty0 439 --
  444, 2004.

\bibitem[Gregory et~al.(2018)Gregory, Lahiri, and Nordman]{gregory2018smooth}
Karl~B Gregory, Soumendra~N Lahiri, and Daniel~J Nordman.
\newblock A smooth block bootstrap for quantile regression with time series.
\newblock \emph{Annals of Statistics}, 46\penalty0 (3):\penalty0 1138--1166,
  2018.

\bibitem[Guttikunda et~al.(2019)Guttikunda, Nishadh, and
  Jawahar]{guttikunda2019air}
Sarath~K Guttikunda, KA~Nishadh, and Puja Jawahar.
\newblock {Air pollution knowledge assessments (APnA) for 20 Indian cities}.
\newblock \emph{Urban Climate}, 27:\penalty0 124--141, 2019.
\newblock \doi{10.1016/j.uclim.2018.11.005}.

\bibitem[Henry et~al.(2002)Henry, Chang, and Spiegelman]{henry2002locating}
Ronald~C Henry, Yu-Shuo Chang, and Clifford~H Spiegelman.
\newblock Locating nearby sources of air pollution by nonparametric regression
  of atmospheric concentrations on wind direction.
\newblock \emph{Atmospheric Environment}, 36\penalty0 (13):\penalty0
  2237--2244, 2002.

\bibitem[Honda(2000)]{honda2000nonparametric}
Toshio Honda.
\newblock Nonparametric estimation of a conditional quantile for
  $\alpha$-mixing processes.
\newblock \emph{Annals of the Institute of Statistical Mathematics},
  52\penalty0 (3):\penalty0 459--470, 2000.

\bibitem[Honda(2013)]{honda2013nonparametric}
Toshio Honda.
\newblock Nonparametric quantile regression with heavy-tailed and strongly
  dependent errors.
\newblock \emph{Annals of the Institute of Statistical Mathematics},
  65\penalty0 (1):\penalty0 23--47, 2013.

\bibitem[Huang et~al.(2017)Huang, Zhang, Chen, and He]{huang2017quantile}
Q~Huang, H~Zhang, J~Chen, and MJJBB He.
\newblock Quantile regression models and their applications: a review.
\newblock \emph{Journal of Biometrics \& Biostatistics}, 8\penalty0
  (10.4172):\penalty0 2155--6180, 2017.

\bibitem[Jagger and Elsner(2006)]{jagger2006climatology}
Thomas~H Jagger and James~B Elsner.
\newblock Climatology models for extreme hurricane winds near the united
  states.
\newblock \emph{Journal of Climate}, 19\penalty0 (13):\penalty0 3220--3236,
  2006.

\bibitem[Jagger and Elsner(2009)]{jagger2009modeling}
Thomas~H Jagger and James~B Elsner.
\newblock Modeling tropical cyclone intensity with quantile regression.
\newblock \emph{International Journal of Climatology: A Journal of the Royal
  Meteorological Society}, 29\penalty0 (10):\penalty0 1351--1361, 2009.
\newblock ISSN 1097-0088.
\newblock \doi{10.1002/joc.1804}.

\bibitem[Jana et~al.(2020)Jana, Sengupta, Kundu, Chakraborty, and
  Shaw]{Jana-JASA}
K.~Jana, D.~Sengupta, S.~Kundu, A.~Chakraborty, and P.~Shaw.
\newblock The statistical face of a region under monsoon rainfall in eastern
  india.
\newblock \emph{Journal of the American Statistical Association}, 115:\penalty0
  1559--1573, 2020.

\bibitem[Jana and Sengupta(2019)]{jana2019improving}
Kaushik Jana and Debasis Sengupta.
\newblock Improving linear quantile regression for replicated data.
\newblock \emph{arXiv preprint arXiv:1901.05369}, 2019.

\bibitem[Jerrett et~al.(2005)Jerrett, Burnett, Ma, Pope~III, Krewski, Newbold,
  Thurston, Shi, Finkelstein, Calle, and Thun]{jerrett2005spatial}
Michael Jerrett, Richard~T Burnett, Renjun Ma, C~Arden Pope~III, Daniel
  Krewski, K~Bruce Newbold, George Thurston, Yuanli Shi, Norm Finkelstein,
  Eugenia~E Calle, and Michael~J Thun.
\newblock Spatial analysis of air pollution and mortality in los angeles.
\newblock \emph{Epidemiology}, pages 727--736, 2005.

\bibitem[Kang et~al.(2019)Kang, Kim, and Elsner]{kang2019contribution}
Nam-Young Kang, Dongjin Kim, and James~B Elsner.
\newblock The contribution of super typhoons to tropical cyclone activity in
  response to enso.
\newblock \emph{Scientific reports}, 9\penalty0 (1):\penalty0 1--6, 2019.

\bibitem[Knight(2001)]{Knight}
Keith Knight.
\newblock {Comparing conditional quantile estimators: First and second order
  considerations}.
\newblock Technical report, Citeseer, 2001.
\newblock URL
  \url{http://citeseerx.ist.psu.edu/viewdoc/download?doi=10.1.1.26.1524&rep=rep1&type=pdf}.
\newblock University of Toronto working paper.

\bibitem[Koenker(2005)]{koenker2005}
Roger Koenker.
\newblock \emph{Quantile regression}.
\newblock Cambridge University press, 2005.

\bibitem[Koenker(2019)]{quantreg}
Roger Koenker.
\newblock \emph{quantreg: Quantile Regression}, 2019.
\newblock URL \url{https://CRAN.R-project.org/package=quantreg}.
\newblock R package version 5.54.

\bibitem[Koenker and Bassett~Jr(1978)]{koenker1978regression}
Roger Koenker and Gilbert Bassett~Jr.
\newblock Regression quantiles.
\newblock \emph{Econometrica: Journal of the Econometric Society}, 46\penalty0
  (1):\penalty0 33--50, 1978.

\bibitem[Kossin et~al.(2013)Kossin, Olander, and Knapp]{Kossin_2013}
James~P. Kossin, Timothy~L. Olander, and Kenneth~R. Knapp.
\newblock Trend analysis with a new global record of tropical cyclone
  intensity.
\newblock \emph{Journal of Climate}, 26\penalty0 (24):\penalty0 9960--9976,
  2013.

\bibitem[Lanzante(1996)]{lanzante1996resistant}
John~R Lanzante.
\newblock Resistant, robust and non-parametric techniques for the analysis of
  climate data: Theory and examples, including applications to historical
  radiosonde station data.
\newblock \emph{International Journal of Climatology: A Journal of the Royal
  Meteorological Society}, 16\penalty0 (11):\penalty0 1197--1226, 1996.

\bibitem[Leider(2012)]{leider2012quantile}
Julien Leider.
\newblock A quantile regression study of climate change in chicago, 1960-2010.
\newblock \emph{Department of Mathematics, Statistics and Computer Science,
  University of Illinois, Chicago}, 2012.

\bibitem[Li et~al.(2020)Li, Li, and Li]{li2020nonparametric}
Degui Li, Qi~Li, and Zheng Li.
\newblock Nonparametric quantile regression estimation with mixed discrete and
  continuous data.
\newblock \emph{Journal of Business \& Economic Statistics}, pages 1--16, 2020.

\bibitem[{London \& Partners}(2021)]{dealroom}
{London \& Partners}.
\newblock {London: Europe's global tech city}.
\newblock \emph{DealRoom.co.}, 2021.
\newblock URL
  \url{https://dealroom.co/uploaded/2021/04/dealroom-london-jan-21.pdf}.

\bibitem[Lu and Fan(2015)]{lu2015weighted}
Xiaoming Lu and Zhaozhi Fan.
\newblock Weighted quantile regression for longitudinal data.
\newblock \emph{Computational Statistics}, 30\penalty0 (2):\penalty0 569--592,
  2015.
\newblock \doi{10.1007/s00180-014-0550-x}.

\bibitem[Nicholls and Kariko(1993)]{Nicholls1993}
Neville Nicholls and Alex Kariko.
\newblock {East Australian Rainfall Events: Interannual Variations, Trends, and
  Relationships with the Southern Oscillation}.
\newblock \emph{Journal of Climate}, 6\penalty0 (6):\penalty0 1141--1152, 1993.
\newblock \doi{10.1175/1520-0442(1993)006<1141:EAREIV>2.0.CO;2}.

\bibitem[Nicholls et~al.(1998)Nicholls, Landsea, and Gill]{nicholls1998recent}
Neville Nicholls, Chris Landsea, and Jon Gill.
\newblock Recent trends in australian region tropical cyclone activity.
\newblock \emph{Meteorology and Atmospheric Physics}, 65\penalty0 (3):\penalty0
  197--205, 1998.

\bibitem[Redden et~al.(2004)Redden, Fernández, and Allison]{Redden_2004}
David~T. Redden, José~R. Fernández, and David~B. Allison.
\newblock A simple significance test for quantile regression.
\newblock \emph{Statistics in Medicine}, 23\penalty0 (16):\penalty0 2587--2597,
  2004.
\newblock ISSN 1097-0258.

\bibitem[Rivas and Gonzalo(2020)]{rivas2020trends}
Mar{\'\i}a Dolores~Gadea Rivas and Jes{\'u}s Gonzalo.
\newblock Trends in distributional characteristics: Existence of global
  warming.
\newblock \emph{Journal of Econometrics}, 214\penalty0 (1):\penalty0 153--174,
  2020.

\bibitem[Ropelewski and Bell(2008)]{Ropelewski2008}
C.~F. Ropelewski and M.~A. Bell.
\newblock {Shifts in the Statistics of Daily Rainfall in South America
  Conditional on ENSO Phase}.
\newblock \emph{Journal of Climate}, 21\penalty0 (5):\penalty0 849--865, 2008.
\newblock \doi{10.1175/2007JCLI1617.1}.

\bibitem[Smith et~al.(2003)Smith, Kolenikov, and Cox]{smith2003spatiotemporal}
Richard~L Smith, Stanislav Kolenikov, and Lawrence~H Cox.
\newblock Spatiotemporal modeling of pm2. 5 data with missing values.
\newblock \emph{Journal of Geophysical Research: Atmospheres}, 108\penalty0
  (D24), 2003.
\newblock \doi{10.1029/2002JD002914}.

\bibitem[Thurston et~al.(2016)Thurston, Burnett, Turner, Shi, Krewski, Lall,
  Ito, Jerrett, Gapstur, Diver, and Pope~III]{thurston2016ischemic}
George~D Thurston, Richard~T Burnett, Michelle~C Turner, Yuanli Shi, Daniel
  Krewski, Ramona Lall, Kazuhiko Ito, Michael Jerrett, Susan~M Gapstur, W~Ryan
  Diver, and C~Arden Pope~III.
\newblock Ischemic heart disease mortality and long-term exposure to
  source-related components of us fine particle air pollution.
\newblock \emph{Environmental health perspectives}, 124\penalty0 (6):\penalty0
  785--794, 2016.
\newblock \doi{10.1289/ehp.1509777}.

\bibitem[Torres et~al.(2005)Torres, Garcia, De~Blas, and
  De~Francisco]{torres2005forecast}
Jose~Luis Torres, Almudena Garcia, Marian De~Blas, and Adolfo De~Francisco.
\newblock Forecast of hourly average wind speed with arma models in navarre
  (spain).
\newblock \emph{Solar energy}, 79\penalty0 (1):\penalty0 65--77, 2005.

\bibitem[Vasseur and Aznarte(2021)]{vasseur2021comparing}
Sebastien~P{\'e}rez Vasseur and Jos{\'e}~L Aznarte.
\newblock Comparing quantile regression methods for probabilistic forecasting
  of no2 pollution levels.
\newblock \emph{Scientific Reports}, 11\penalty0 (1):\penalty0 1--8, 2021.

\bibitem[Wahba(1990)]{wahba1990spline}
Grace Wahba.
\newblock \emph{Spline models for observational data}.
\newblock SIAM, 1990.

\bibitem[Watson-Parris(2021)]{Watson-Parris2021}
D.~Watson-Parris.
\newblock Machine learning for weather and climate are worlds apart.
\newblock \emph{Phil. Trans. R. Soc. A.}, 379, 2021.
\newblock ISSN 1364-503X.
\newblock \doi{https://doi.org/10.1098/rsta.2020.0098}.

\bibitem[Wei et~al.(2019)Wei, Kehm, Goldberg, and Terry]{wei2019applications}
Ying Wei, Rebecca~D Kehm, Mandy Goldberg, and Mary~Beth Terry.
\newblock Applications for quantile regression in epidemiology.
\newblock \emph{Current Epidemiology Reports}, 6\penalty0 (2):\penalty0
  191--199, 2019.

\bibitem[Wu(2005)]{wu2005nonlinear}
Wei~Biao Wu.
\newblock Nonlinear system theory: Another look at dependence.
\newblock \emph{Proceedings of the National Academy of Sciences}, 102\penalty0
  (40):\penalty0 14150--14154, 2005.

\bibitem[Wu and Zhao(2007)]{wu2007inference}
Wei~Biao Wu and Zhibiao Zhao.
\newblock Inference of trends in time series.
\newblock \emph{Journal of the Royal Statistical Society: Series B (Statistical
  Methodology)}, 69\penalty0 (3):\penalty0 391--410, 2007.

\bibitem[Ying et~al.(2011)Ying, Chen, and Wu]{ying2011climate}
Ming Ying, Baode Chen, and Guoxiong Wu.
\newblock Climate trends in tropical cyclone-induced wind and precipitation
  over mainland china.
\newblock \emph{Geophysical Research Letters}, 38\penalty0 (1), 2011.

\bibitem[Zhao and Wu(2006)]{zhao2006kernel}
Zhibiao Zhao and Wei~Biao Wu.
\newblock Kernel quantile regression for nonlinear stochastic models.
\newblock \emph{Technical ReportNo}, 572, 2006.

\bibitem[Zhao and Wu(2008)]{zhao2008confidence}
Zhibiao Zhao and Wei~Biao Wu.
\newblock Confidence bands in nonparametric time series regression.
\newblock \emph{The Annals of Statistics}, 36\penalty0 (4):\penalty0
  1854--1878, 2008.

\bibitem[Ziegelmann(2005)]{ziegelmann2005nonparametric}
Flavio~A Ziegelmann.
\newblock A nonparametric least-absolute-deviations estimator of volatility
  functions.
\newblock \emph{XXVII Encontro Brasileiro de Econometria, Natal}, 10, 2005.

\end{thebibliography}

\section{Proofs}
\label{sec:proof}

\begin{proof}[Proof of \Cref{thm:clt}]
Note that
\begin{equation}
\label{eq:main-equation}
\frac{\sqrt{nb_n}\sqrt{\hat g(x;b_n)}}{\sigma(x)\sqrt{\phi_K}}  \left[ \hat\mu(x; b_n) - \mu(x) \right] = \frac{1}{\sigma(x)\sqrt{\phi_K nb_n \hat g(x;b_n)}} \sum_{t=1}^n K\left(\frac{x-\x_t}{b_n}\right) \{\mu(\x_t) - \mu(x) + \sigma(\x_t)e_t\}
\end{equation}

Let $r(x;b_n) = g(x)/\hat g(x;b_n)$ and 
\begin{align}
\label{eq:defintion-Sx}
S(x;b_n) &= \frac{1}{\sigma(x)\sqrt{\phi_K nb_n g(x)}} \sum_{t=1}^n K\left(\frac{x-\x_t}{b_n}\right) \sigma(\x_t)e_t, \\
\label{eq:definition-Bx}
B(x;b_n) &= \frac{1}{nb_n g(x)} \sum_{t=1}^n K\left(\frac{x-\x_t}{b_n}\right) \{\mu(\x_t) - \mu(x)\}.
\end{align}

It is easy to observe that \cref{eq:main-equation} can be rewritten as 
\begin{equation}
\label{eq:two-parts}
    \frac{\sqrt{nb_n}\sqrt{\hat g(x;b_n)}}{\sigma(x)\sqrt{\phi_K}} \left[ \hat\mu(x; b_n) - \mu(x) - r(x;b_n)B(x;b_n)\right] = \sqrt{r(x;b_n)}S(x;b_n).
\end{equation}

We shall treat the two terms $S(x;b_n)$ and $B(x;b_n)$ separately. First, $S(x;b_n)$ can be written as the sum of $W_t(x;b_n)$ where $$W_t(x;b_n) = \frac{K\left(\frac{x-\x_t}{b_n}\right) \sigma(\x_t)e_t}{\sigma(x)\sqrt{\phi_K nb_n g(x)}}.$$

Recall \Cref{asmp:independence} and let $\G_t$ be the sigma-field generated by $(\hdots,\eps_{t-1},\eps_t,\eps_{t+1};\hdots,e_{t-1},e_t)$. Then, the sequence $\{W_t(x;b_n)\}_{t=1}^n$ form a martingale difference sequence with respect to the filtration $\G_t$. Next, denoting $c(x)=(\phi_Kg(x)\sigma^2(x))^{-1}$, in view of \Cref{asmp:regressor-process} and the above results, we can write
\begin{eqnarray*}
\sum_{t=1}^n \E\left(W_t^2(x;b_n) \given \G_{t-1}\right) &=& \frac{c(x)}{nb_n} \sum_{t=1}^n \E\left[ K^2\left(\frac{x-\x_t}{b_n}\right) \sigma^2(\x_t)e_t^2 \given \G_{t-1} \right] \\
&=& \frac{c(x)}{nb_n} \sum_{t=1}^n K^2\left(\frac{x-\x_t}{b_n}\right) \sigma^2(\x_t).
\end{eqnarray*}

Let $u_t = K^2((x-\x_t)/b_n)\sigma^2(\x_t)$. It is easy to work out that, as $n\to\infty$,
\begin{equation}
\label{eq:main-proof-step1}
\frac{1}{nb_n}\sum_{t=1}^n \E(u_t) \to \phi_Kg(x)\sigma^2(x).
\end{equation}

Further, let $u_t-\E(u_t) = v_t + w_t$, where $v_t = u_t-\E(u_t\given \F_{t-1})$, $w_t= \E(u_t\given \F_{t-1}) - \E(u_t)$. Observe that $\{v_t\}_{t=1}^n$ form a martingale difference sequence with respect to the filtration $\F_t$ and $\E(v_t^2) = O(b_n)$. Thus,
\begin{equation}
\label{eq:main-proof-step2}
\frac{1}{nb_n}\sum_{t=1}^n v_t \to 0.
\end{equation}

On the other hand, $w_t = \int K^2(u) \sigma^2(x-ub_n) \{g(x-ub_n\given \F_{t-1}) - g(x-ub_n)\}\ du$. Then, using Lemma 1 from \cite{zhao2008confidence}, it can be shown that
\begin{equation}
\label{eq:main-proof-step3}
\frac{1}{nb_n} \norm{\sum_{t=1}^n w_t} \to 0.
\end{equation}

Combining equations (\ref{eq:main-proof-step1}), (\ref{eq:main-proof-step2}) and (\ref{eq:main-proof-step3}), 
\begin{equation}
\label{eq:main-proof-step4}
\sum_{t=1}^n \E\left(W_t^2(x;b_n) \given \G_{t-1}\right) = \frac{c(x)}{nb_n} \sum_{t=1}^n K^2\left(\frac{x-\x_t}{b_n}\right) \sigma^2(\x_t) \to 1.
\end{equation}

Next step is to look at the Lindeberg condition for $W_t(x;b_n)$.
\begin{eqnarray*}
\sum_{t=1}^n \E[W_t^2(x;b_n)\ \ind\{\abs{W_t(x;b_n)} \geqslant \delta\}] &=& \frac{c(x)}{nb_n} \sum_{t=1}^n \E\left[u_te_t^2 \ \ind\left\{ \abs{u_te_t^2} \geqslant \delta^2nb_n/c(x) \right\}\right] \\
&=& \frac{c(x)}{b_n} \E\left[u_0e_0^2 \ \ind\left\{ \abs{u_0e_0^2} \geqslant \delta^2nb_n/c(x) \right\}\right].
\end{eqnarray*}

Under the aforementioned assumptions, $u_0$ is independent of $e_0$, $\E(u_0) = O(b_n)$ and $\sup \abs{u_0} \leqslant M$ for some finite constant $M$ and for large enough $n$. Thus, $\ind\{ \abs{u_0e_0^2} \geqslant \delta^2nb_n/c(x)\} \leqslant \ind\{ \abs{e_0^2} \geqslant \delta^2nb_n/Mc(x)\}$, which further implies that $\E[e_0^2\ \ind\{ \abs{u_0e_0^2} \geqslant \delta^2nb_n/c(x)\}] \to 0$ as $n \to \infty$. Clearly, as $n\to\infty$, the following Lindeberg condition holds.
\begin{equation}
\label{eq:lindeberg}
\sum_{t=1}^n \E[W_t^2(x;b_n)\ \ind\{\abs{W_t(x;b_n)} \geqslant \delta\}] \to 0.
\end{equation}

Hence, using martingale central limit theorem, we can say that $S(x;b_n) \to \N(0,1)$.

We now turn our attention to $B(x;b_n)$. Keeping the notations same as above for convenience, let $u_t = K((x-\x_t)/b_n)\{\mu(\x_t)-\mu(x)\}$. Similarly as above, we write it as $u_t = v_t + w_t + E(u_t)$ where $v_t = u_t - \E(u_t\given \F_{t-1})$ and $w_t = \E(u_t\given \F_{t-1}) - \E(u_t)$. Once again, results identical to \cref{eq:main-proof-step2} and \cref{eq:main-proof-step3} hold true here as well. These can be proved following similar idea as before and are omitted. Further, simple calculations show that
\begin{equation}
\label{eq:main-proof-bias-expectation}
\frac{1}{nb_n}\sum_{t=1}^n \E(u_t) \to b_n^2\psi_K (\mu\ddash(x)g(x) + 2\mu\dash(x)g\dash(x)),
\end{equation}
as $n\to\infty$. This, along with the previous results, imply that $B(x;b_n)-\delta(x;b_n) \to 0$ in probability.

As a last piece of the proof, we use the fact that $r(x;b_n) \to 1$ for all $x$ as $n\to\infty$. This is a well known result and has been discussed in multiple papers. Combining the asymptotic properties of $S(x;b_n)$, $B(x;b_n)$ and $r(x;b_n)$, we get the desired result.
\end{proof}

\begin{proof}[Proof of \Cref{thm:sigma-consistency}]
Similar to the previous proof, we start from the fact that
\begin{equation}
\label{eq:sigmaproof-step1}
    \hat\sigma^2(x;b_n) = \frac{1}{nb_n\hat g(x;b_n)}\sum_{t=1}^n K_2
    \left(\frac{x-\x_t}{b_n}\right) \left( \mu(\x_t) + \sigma(\x_t)e_t -\hat\mu^*(\x_t;b_n)\right)^2.
\end{equation}

Let us define $S(x;b_n)$ in the same way as in \cref{eq:defintion-Sx}, with $K(\cdot)$ therein replaced by $K_2(\cdot)$. Let
\begin{align}
\label{eq:sigmaproof-S1-defn}
S_1(x;b_n) &= \frac{1}{nb_n \hat g(x;b_n)} \sum_{t=1}^n K_2\left(\frac{x-\x_t}{b_n}\right) \sigma(\x_t)e_t,  \\
\label{eq:sigmaproof-S2-defn}
S_2(x;b_n) &= \frac{1}{nb_n \hat g(x;b_n)} \sum_{t=1}^n K_2\left(\frac{x-\x_t}{b_n}\right) \sigma^2(\x_t)e_t^2.
\end{align}

Then, following the proof of asymptotic normality of $S(x;b_n)$ in \Cref{thm:clt} and using \Cref{corr:jackknife}, it can be argued that uniformly for all $x$,
\begin{equation}
\label{eq:sigmaproof-step2}
    \hat\mu^*(x;b_n) - \mu(x) = D_n + d_n, \; \text{where} \; D_n=S_1(x;b_n)=O((nb_n)^{-1/2}), \; \text{and} \; d_n = o((nb_n)^{-1/2}).
\end{equation}

Combining \cref{eq:sigmaproof-step1} and \cref{eq:sigmaproof-step2}, it is straightforward to note that $\hat\sigma^2(x;b_n)$ is asymptotically equivalent to $S_2(x;b_n)$. Next, in view of $\E(e_t^2)=1$, we can write
\begin{align}
    S_2(x;b_n) - \sigma^2(x) &= \frac{1}{nb_n \hat g(x;b_n)} \sum_{t=1}^n K_2\left(\frac{x-\x_t}{b_n}\right) \left\{\sigma^2(\x_t)e_t^2 - \sigma^2(x)\right\} \nonumber \\
    &= \frac{1}{nb_n \hat g(x;b_n)} \sum_{t=1}^n K_2\left(\frac{x-\x_t}{b_n}\right)\sigma^2(\x_t)\left\{e_t^2 - \E(e_t^2)\right\} + \nonumber \\
    \label{eq:sigmaproof_step3}
    &  \quad \frac{1}{nb_n \hat g(x;b_n)} \sum_{t=1}^n K_2\left(\frac{x-\x_t}{b_n}\right) \left\{\sigma^2(\x_t) - \sigma^2(x)\right\}.
\end{align}

For the first term above, using $a_n=nb_n/\log n$, let us write
\begin{equation}
    \label{eq:lemma-step1}
    \begin{split}
    K_2\left(\frac{x-\x_t}{b_n}\right) \sigma^2(\x_t)\{e_t^2-\E(e_t^2)\} = K_2\left(\frac{x-\x_t}{b_n}\right) \sigma^2(\x_t)\{e_t^2\ind\{e_t^2>a_n\}-\E(e_t^2\ind\{e_t^2>a_n\})\} \\ 
    + K_2\left(\frac{x-\x_t}{b_n}\right) \sigma^2(\x_t)\{e_t^2\ind\{e_t^2\leqslant a_n\}-\E(e_t^2\ind\{e_t^2\leqslant a_n\})\}.
    \end{split}
\end{equation}

Denote the above two terms as $T_{1,t}(x;b_n)$ and $T_{2,t}(x;b_n)$. Let $T_1(x;b_n)=\sum_{t=1}^n T_{1,t}(x;b_n)$ and $T_2(x;b_n)=\sum_{t=1}^n T_{2,t}(x;b_n)$. It is easy to note that $\{T_{1,t}(x;b_n)\}_{t=1}^n$ forms martingale difference sequence with respect to the filtration $\G_t$. Also, $\E[e_t^4\ind\{e_t^2>a_n\}]$ is $O(1)$. Thus, $\norm{T_1(x;b_n)}^2 = O(nb_n)$ and $\norm{\partial T_1(x;b_n)/\partial x}^2 = O(n/b_n)$, thereby implying that $\E[\sup_x\abs{T_1(x;b_n)}^2]=O(n/b_n)$ and that  $\sup_x\abs{T_1(x;b_n)} = O(\sqrt{n/b_n})$.

On the other hand, $\{T_{2,t}(x;b_n)/a_n\}_{t=1}^n$ is a uniformly bounded martingale difference sequence with respect to the filtration $\G_t$. Following Lemma 4 of \cite{zhao2006kernel}, it can be argued that $\sup_x\abs{T_2(x;b_n)} = O(\sqrt{nb_n\log n})$. Using this, in conjunction with the above result for $T_1(x;b_n)$, we get that the first term in \cref{eq:sigmaproof_step3} is $O(\sqrt{1/nb_n^3}+\sqrt{\log n/nb_n})$.

For the second term in \cref{eq:sigmaproof_step3}, recall the expression for $B(x;b_n)$ from \cref{eq:definition-Bx}, and note that in proving the asymptotic result of $B(x;b_n)$, we only used the properties of $\mu(\cdot)$ given by \Cref{asmp:regularity-conditions}, all of which are true for $\sigma^2(\cdot)$ as well. Thus, exactly similar result can be proved if we replace $\mu(\cdot)$ by $\sigma^2(\cdot)$ in that expression. In particular, we can show that
\begin{equation}
    \frac{1}{nb_n \hat g(x;b_n)} \sum_{t=1}^n K_2\left(\frac{x-\x_t}{b_n}\right) \left\{\sigma^2(\x_t) - \sigma^2(x)\right\} - \delta_\sigma(x;b_n) \to 0,
\end{equation}
where $\delta_\sigma(x;b_n) = b_n^2\psi_K(2\sigma(x)\sigma\ddash(x) + 2(\sigma\dash(x))^2 + 4\sigma(x)\sigma\dash(x)g\dash(x)/g(x))$. Now, using the assumptions about the bandwidth function $b_n$, it is easy to see that both terms in \cref{eq:sigmaproof_step3}  converge to 0 as $n\to\infty$ and that concludes our proof.
\end{proof}


\section{Additional simulation results}
\label{sec:additional-results}

In our simulation study, we use four different data generating processes (DGP), as given by \Cref{tab:DGP}. The results corresponding to two of the DGPs are presented in the main paper. Here, we focus on the other two processes. Note that NLQRM refers to the DGP where $\mu(\x_t)$ is nonlinear in $\x_t$ and $\sigma(\x_t)$ is constant whereas NLHQRM denotes the process with nonlinear $\mu(\x_t)$ and non-constant $\sigma(\x_t)$. Recall that we run 1000 simulations for every DGP and for every such experiment, the root mean squared error (RMSE) of the fitted values, that is the square root of the mean of $(\hat\mu(x)-\mu(x))^2$ over all $x$ in the training set, is computed. The mean RMSE corresponding to NLQRM for the three candidate models for different cases are presented in \Cref{tab:MSE-NLQRM} and the same for NLHQRM are displayed in \Cref{tab:MSE-NLHQRM}. In all the following results, KB corresponds to the method proposed by \cite{koenker1978regression} (refer to \cref{eq:estimate-kb} in the main paper), JS is the approach proposed by \cite{jana2019improving} (refer to \cref{eq:estimate-js} in the main paper), and NP stands for our proposed approach (see \cref{eq:mu-estimate} in the main paper).

\begin{table}[!hbt]
\centering
\caption{Empirical RMSE (mean taken over 1000 simulations) of fitted quantiles using the three different approaches for different values of $\tau$, $K$ and $n$, when the data are generated from a NLQRM.}
\label{tab:MSE-NLQRM}
\begin{tabular}{|cc|ccc|ccc|ccc|}
    \hline
    & & \multicolumn{3}{c|}{$n$=30} & \multicolumn{3}{c|}{$n$=50} & \multicolumn{3}{c|}{$n$=100} \\
	$\tau$ & Method & $K$=50 & $K$=100 & $K$=500 & $K$=50 & $K$=100 & $K$=500 & $K$=50 & $K$=100 & $K$=500 \\
	\hline
	0.7 & KB & 1.033 & 1.026 & 1.026 & 1.048 & 1.029 & 1.022 & 1.044 & 1.041 & 1.039 \\ 
   & JS & 0.995 & 0.989 & 0.987 & 1.009 & 0.992 & 0.986 & 1.006 & 1.004 & 1.001 \\ 
   & NP & 0.184 & 0.140 & 0.099 & 0.189 & 0.144 & 0.101 & 0.193 & 0.148 & 0.104 \\ 
   \hline
  0.8 & KB & 1.088 & 1.069 & 1.061 & 1.085 & 1.089 & 1.069 & 1.096 & 1.090 & 1.077 \\ 
   & JS & 1.003 & 0.988 & 0.980 & 1.003 & 1.006 & 0.990 & 1.011 & 1.007 & 0.998 \\ 
   & NP & 0.197 & 0.149 & 0.103 & 0.203 & 0.152 & 0.106 & 0.206 & 0.157 & 0.109 \\ 
   \hline
  0.9 & KB & 1.185 & 1.171 & 1.151 & 1.188 & 1.178 & 1.155 & 1.204 & 1.187 & 1.169 \\ 
   & JS & 1.010 & 0.998 & 0.984 & 1.010 & 1.006 & 0.990 & 1.019 & 1.012 & 1.003 \\ 
   & NP & 0.228 & 0.171 & 0.114 & 0.234 & 0.176 & 0.117 & 0.236 & 0.179 & 0.120 \\ 
   \hline
  0.95 & KB & 1.309 & 1.258 & 1.245 & 1.320 & 1.283 & 1.251 & 1.314 & 1.291 & 1.264 \\ 
   & JS & 1.029 & 0.998 & 0.988 & 1.032 & 1.011 & 0.996 & 1.025 & 1.016 & 1.003 \\ 
   & NP & 0.261 & 0.202 & 0.130 & 0.268 & 0.206 & 0.135 & 0.273 & 0.212 & 0.139 \\
	\hline
\end{tabular}
\end{table}

\begin{table}[!hbt]
\centering
\caption{Empirical RMSE (mean taken over 1000 simulations) of fitted quantiles using the three different approaches for different values of $\tau$, $K$ and $n$, when the data are generated from a NLHQRM.}
\label{tab:MSE-NLHQRM}
\begin{tabular}{|cc|ccc|ccc|ccc|}
    \hline
    & & \multicolumn{3}{c|}{$n$=30} & \multicolumn{3}{c|}{$n$=50} & \multicolumn{3}{c|}{$n$=100} \\
	$\tau$ & Method & $K$=50 & $K$=100 & $K$=500 & $K$=50 & $K$=100 & $K$=500 & $K$=50 & $K$=100 & $K$=500 \\
	\hline
	0.7 & KB & 1.062 & 1.045 & 1.030 & 1.065 & 1.047 & 1.034 & 1.075 & 1.062 & 1.042 \\ 
   & JS & 1.047 & 1.032 & 1.017 & 1.049 & 1.033 & 1.022 & 1.059 & 1.047 & 1.030 \\ 
   & NP & 0.254 & 0.185 & 0.122 & 0.257 & 0.191 & 0.124 & 0.263 & 0.195 & 0.129 \\ 
   \hline
  0.8 & KB & 1.100 & 1.082 & 1.064 & 1.111 & 1.086 & 1.072 & 1.118 & 1.100 & 1.082 \\ 
   & JS & 1.044 & 1.030 & 1.013 & 1.052 & 1.032 & 1.022 & 1.057 & 1.044 & 1.032 \\ 
   & NP & 0.270 & 0.198 & 0.128 & 0.278 & 0.205 & 0.132 & 0.282 & 0.208 & 0.136 \\ 
   \hline
  0.9 & KB & 1.219 & 1.174 & 1.142 & 1.219 & 1.189 & 1.156 & 1.227 & 1.185 & 1.158 \\ 
   & JS & 1.066 & 1.037 & 1.014 & 1.068 & 1.051 & 1.027 & 1.070 & 1.046 & 1.030 \\ 
   & NP & 0.312 & 0.233 & 0.146 & 0.323 & 0.238 & 0.152 & 0.331 & 0.242 & 0.154 \\ 
   \hline
  0.95 & KB & 1.359 & 1.285 & 1.250 & 1.346 & 1.292 & 1.251 & 1.354 & 1.305 & 1.263 \\ 
   & JS & 1.097 & 1.046 & 1.029 & 1.088 & 1.055 & 1.033 & 1.084 & 1.058 & 1.040 \\ 
   & NP & 0.360 & 0.277 & 0.174 & 0.374 & 0.286 & 0.178 & 0.380 & 0.291 & 0.182 \\ 
	\hline
\end{tabular}
\end{table}

In both of these tables, we can see that the NP method records better RMSE than KB or JS methods. For the proposed nonparametric approach, the RMSE decreases with more replicates, but the values are more or less stable for different values of $n$. For our method, this is in line with what we observed for SQRM or GQRM (refer to the main paper). Interestingly, KB or JS do not display similar behavior and the RMSE values are always similar. It is clear that for a nonlinear mean function, these two approaches cannot perform well. It is also worth noting that for higher quantile ($\tau=0.95$), the error increases marginally for all of the methods. 

Next, we look at the prediction RMSE in different cases. Results for NLQRM are displayed in \Cref{tab:predMSE-NLQRM} and the same for NLHQRM are shown in \Cref{tab:predMSE-NLHQRM}. Once again, we observe that the errors are less for our proposed nonparametric approach than the other two methods. The difference is more for the lower quantiles and is less for the 95\% quantile.

\begin{table}[!hbt]
\centering
\caption{Prediction RMSE (mean taken over 1000 simulations) for the three different approaches for different values of $\tau$, $K$ and $n$, when the data are generated from a NLQRM.}
\label{tab:predMSE-NLQRM}
\begin{tabular}{|cc|ccc|ccc|ccc|}
    \hline
    & & \multicolumn{3}{c|}{$n$=30} & \multicolumn{3}{c|}{$n$=50} & \multicolumn{3}{c|}{$n$=100} \\
	$\tau$ & Method & $K$=50 & $K$=100 & $K$=500 & $K$=50 & $K$=100 & $K$=500 & $K$=50 & $K$=100 & $K$=500 \\
	\hline
	0.7 & KB & 1.085 & 1.085 & 1.074 & 1.070 & 1.061 & 1.048 & 1.069 & 1.065 & 1.047 \\ 
   & JS & 1.055 & 1.042 & 1.044 & 1.028 & 1.029 & 1.015 & 1.029 & 1.028 & 1.012 \\ 
   & NP & 0.223 & 0.175 & 0.136 & 0.211 & 0.167 & 0.123 & 0.208 & 0.161 & 0.117 \\ 
   \hline
  0.8 & KB & 1.143 & 1.106 & 1.116 & 1.111 & 1.103 & 1.112 & 1.107 & 1.098 & 1.091 \\ 
   & JS & 1.053 & 1.034 & 1.038 & 1.036 & 1.030 & 1.029 & 1.024 & 1.016 & 1.015 \\ 
   & NP & 0.234 & 0.184 & 0.139 & 0.228 & 0.177 & 0.132 & 0.221 & 0.170 & 0.123 \\ 
   \hline
  0.9 & KB & 1.222 & 1.200 & 1.202 & 1.219 & 1.199 & 1.186 & 1.213 & 1.193 & 1.171 \\ 
   & JS & 1.058 & 1.033 & 1.031 & 1.039 & 1.028 & 1.017 & 1.034 & 1.022 & 1.011 \\ 
   & NP & 0.270 & 0.205 & 0.158 & 0.259 & 0.200 & 0.143 & 0.252 & 0.190 & 0.135 \\ 
   \hline
  0.95 & KB & 1.327 & 1.301 & 1.278 & 1.346 & 1.299 & 1.270 & 1.327 & 1.298 & 1.282 \\ 
   & JS & 1.068 & 1.044 & 1.022 & 1.060 & 1.035 & 1.021 & 1.037 & 1.022 & 1.020 \\ 
   & NP & 0.308 & 0.237 & 0.161 & 0.297 & 0.230 & 0.157 & 0.288 & 0.224 & 0.153 \\
	\hline
\end{tabular}
\end{table}

\begin{table}[!hbt]
\centering
\caption{Prediction RMSE (mean taken over 1000 simulations) for the three different approaches for different values of $\tau$, $K$ and $n$, when the data are generated from a NLHQRM.}
\label{tab:predMSE-NLHQRM}
\begin{tabular}{|cc|ccc|ccc|ccc|}
    \hline
    & & \multicolumn{3}{c|}{$n$=30} & \multicolumn{3}{c|}{$n$=50} & \multicolumn{3}{c|}{$n$=100} \\
	$\tau$ & Method & $K$=50 & $K$=100 & $K$=500 & $K$=50 & $K$=100 & $K$=500 & $K$=50 & $K$=100 & $K$=500 \\
	\hline
	0.7 & KB & 1.113 & 1.069 & 1.097 & 1.090 & 1.098 & 1.062 & 1.074 & 1.068 & 1.059 \\ 
   & JS & 1.101 & 1.058 & 1.085 & 1.074 & 1.084 & 1.051 & 1.059 & 1.053 & 1.047 \\ 
   & NP & 0.289 & 0.221 & 0.159 & 0.287 & 0.218 & 0.145 & 0.275 & 0.206 & 0.142 \\ 
   \hline
  0.8 & KB & 1.168 & 1.140 & 1.124 & 1.152 & 1.126 & 1.100 & 1.124 & 1.103 & 1.120 \\ 
   & JS & 1.115 & 1.088 & 1.073 & 1.092 & 1.074 & 1.053 & 1.065 & 1.049 & 1.068 \\ 
   & NP & 0.319 & 0.234 & 0.167 & 0.304 & 0.229 & 0.153 & 0.296 & 0.221 & 0.152 \\ 
   \hline
  0.9 & KB & 1.261 & 1.225 & 1.191 & 1.234 & 1.208 & 1.183 & 1.236 & 1.202 & 1.174 \\ 
   & JS & 1.113 & 1.092 & 1.062 & 1.090 & 1.080 & 1.059 & 1.084 & 1.061 & 1.047 \\ 
   & NP & 0.360 & 0.272 & 0.182 & 0.352 & 0.263 & 0.172 & 0.343 & 0.258 & 0.167 \\ 
   \hline
  0.95 & KB & 1.378 & 1.328 & 1.287 & 1.375 & 1.324 & 1.277 & 1.378 & 1.310 & 1.271 \\ 
   & JS & 1.129 & 1.099 & 1.077 & 1.120 & 1.083 & 1.061 & 1.106 & 1.072 & 1.054 \\ 
   & NP & 0.408 & 0.322 & 0.210 & 0.404 & 0.314 & 0.199 & 0.399 & 0.305 & 0.195 \\
	\hline
\end{tabular}
\end{table}

\begin{figure}[!hbt]
    \centering
 \caption{Prediction MSE (mean taken over 1000 simulations) corresponding to different future time-points for the three different approaches. Data are generated from a SQRM with $n=100$, and the results are shown for different values of $\tau$ and $K$.}
 \label{fig:predMSE-SQRM}
    \includegraphics[width = \textwidth,keepaspectratio]{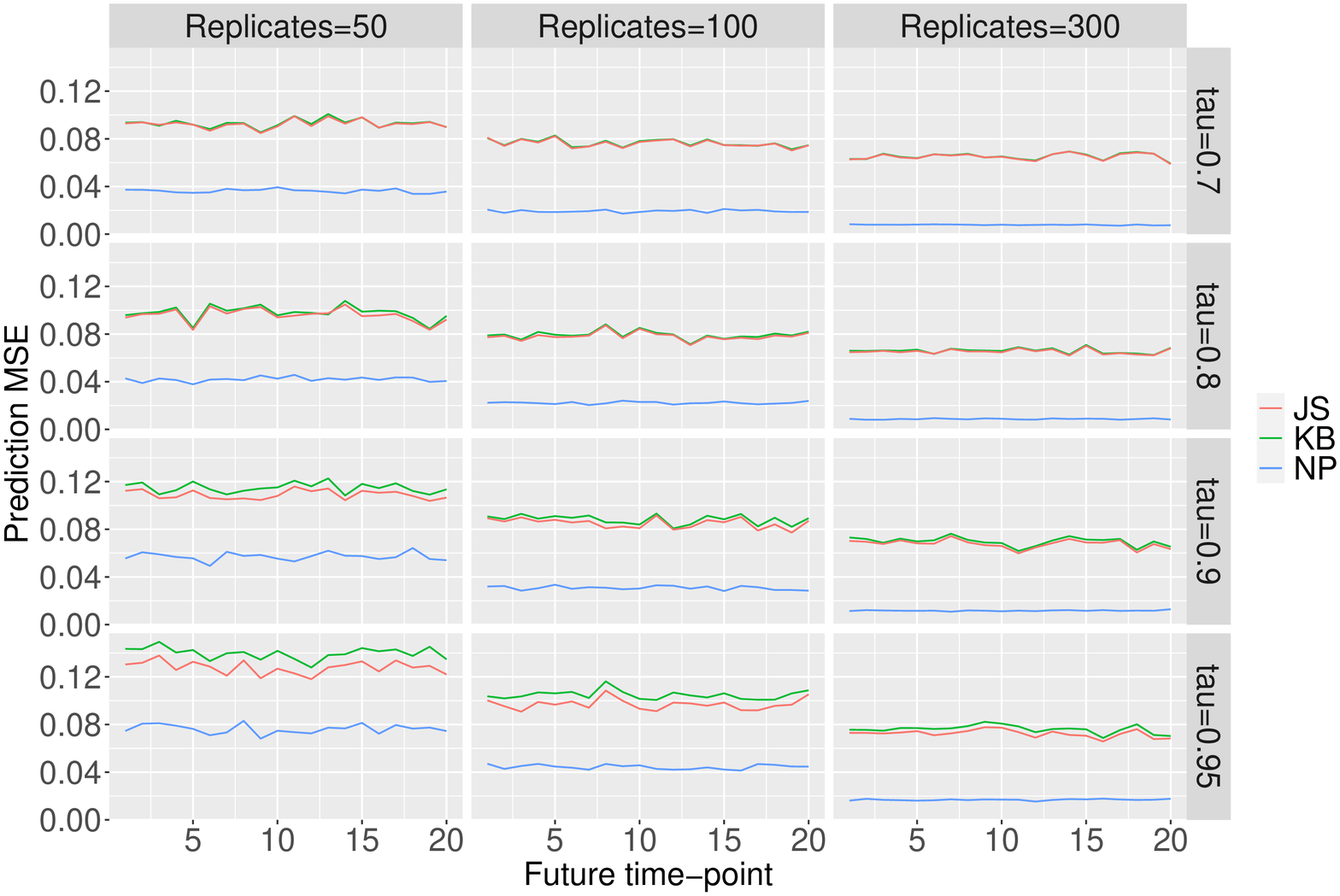}
\end{figure}

\begin{figure}[!hbt]
    \centering
    \includegraphics[width = \textwidth,keepaspectratio]{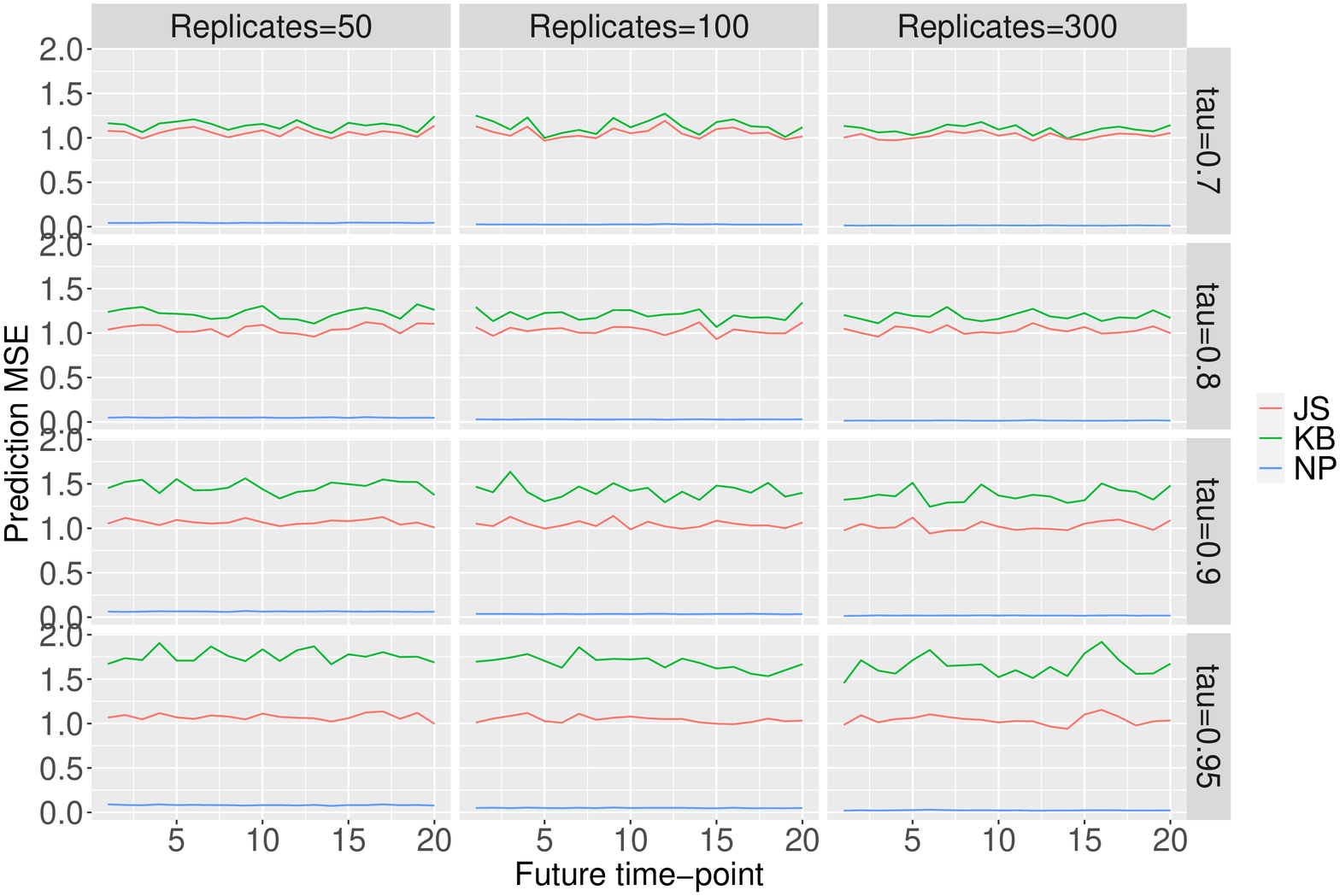}
    \caption{Prediction MSE (mean taken over 1000 simulations) corresponding to different future time-points for the three different approaches. Data are generated from a NLQRM with $n=100$, and the results are shown for different values of $\tau$ and $K$.}
    \label{fig:predMSE-NLQRM}
\end{figure}

\begin{figure}[!hbt]
    \centering
    \includegraphics[width = \textwidth,keepaspectratio]{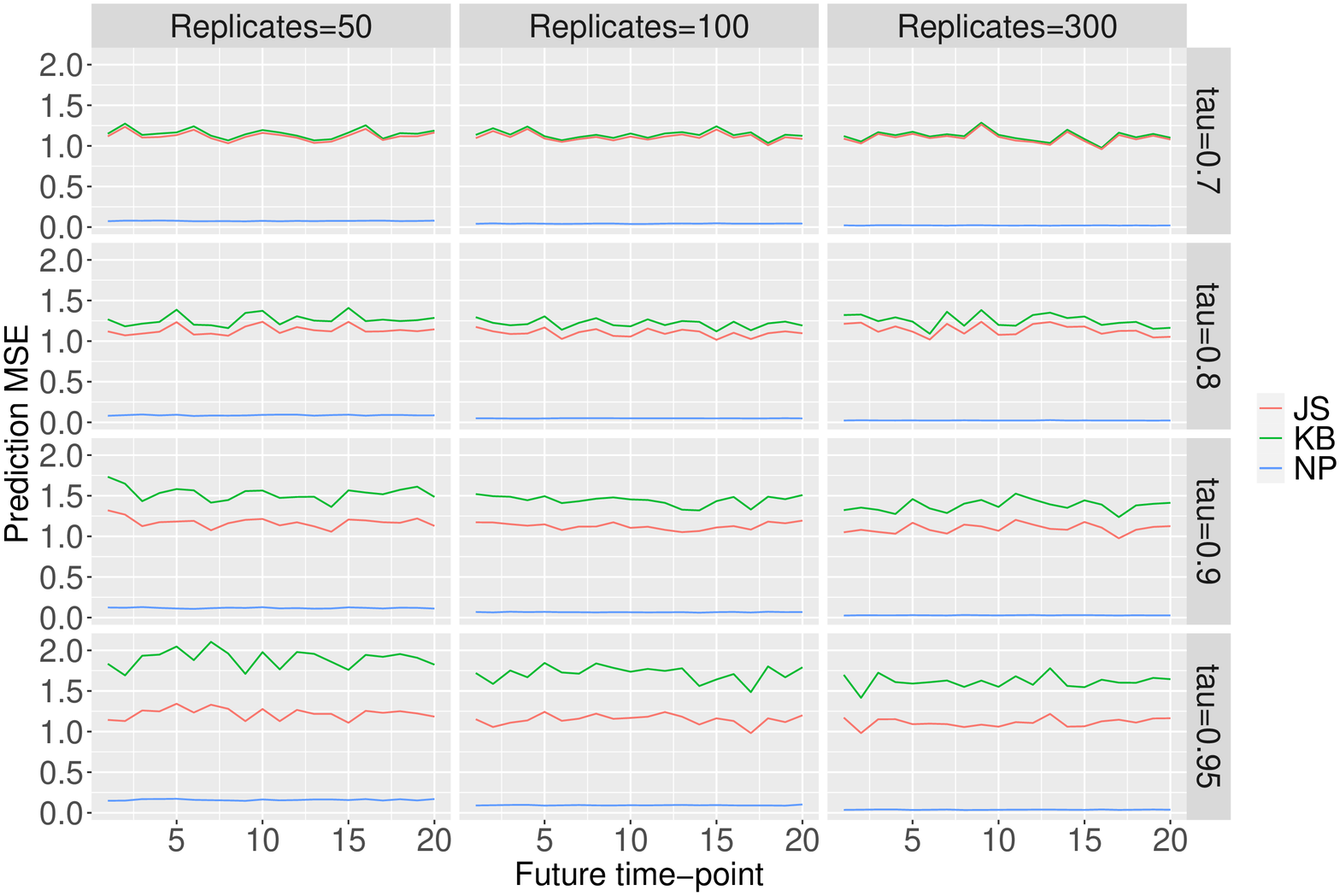}
    \caption{Prediction MSE (mean taken over 1000 simulations) corresponding to different future time-points for the three different approaches. Data are generated from a NLHQRM with $n=100$, and the results are shown for different values of $\tau$ and $K$.}
    \label{fig:predMSE-NLHQRM}
\end{figure}

\begin{figure}[!hbt]
    \centering
    \caption{Prediction MSE (mean taken over 1000 simulations) corresponding to different future time-points for the three different approaches. Data are generated from a GQRM with $n=100$, and the results are shown for different values of $\tau$ and $K$.}
\label{fig:predMSE-GQRM}
    \includegraphics[width = \textwidth,keepaspectratio]{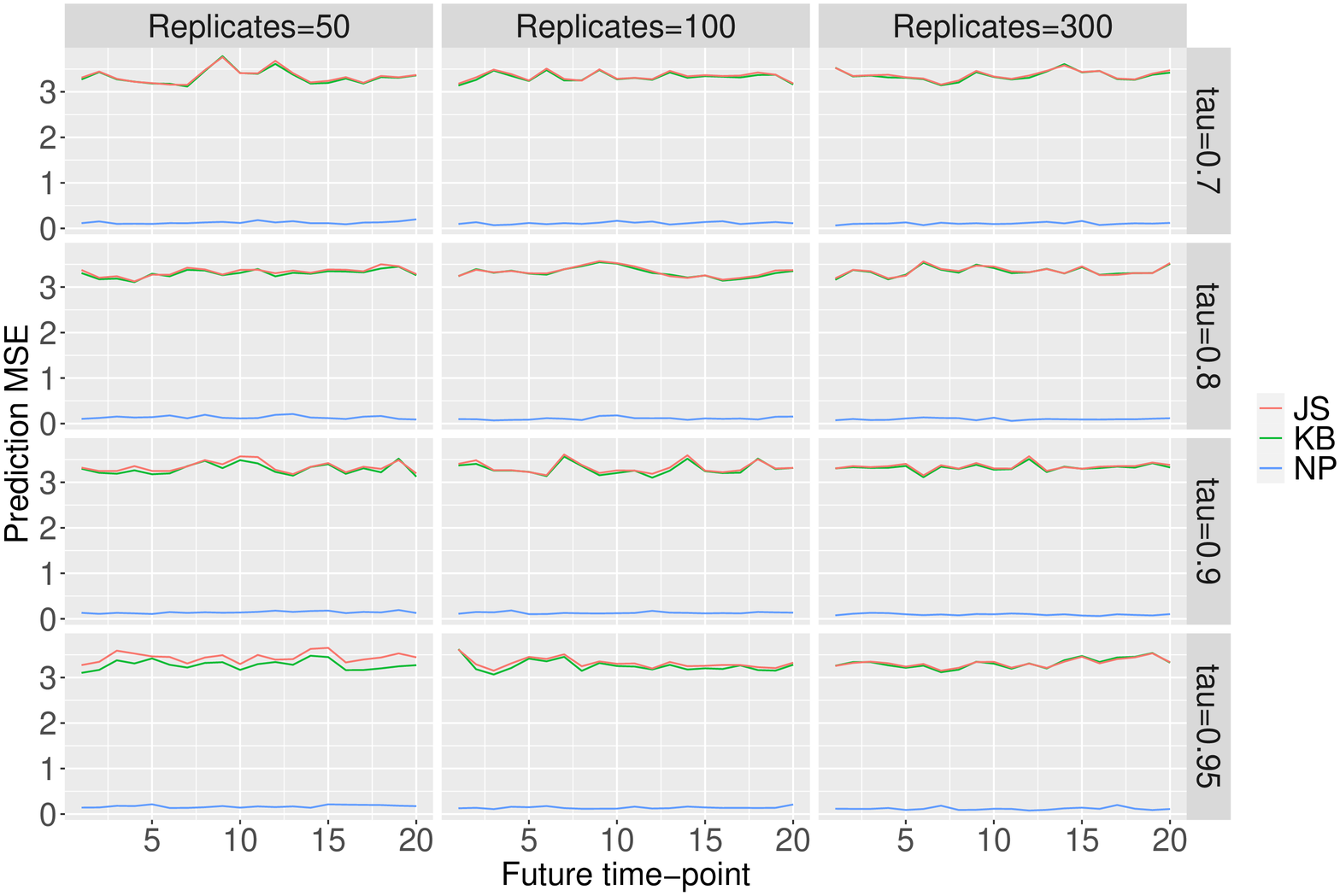}
\end{figure}

As a final piece in this analysis, to better understand the efficiency of the proposed approach, we focus on $n=100$, and take a look at the prediction MSE for all future time-points (1-step ahead to 20-steps ahead). These results, for different values of $K$ and $\tau$, are plotted in \Cref{fig:predMSE-SQRM} (for SQRM), \Cref{fig:predMSE-NLQRM} (for NLQRM), \Cref{fig:predMSE-NLHQRM} (for NLHQRM) and \Cref{fig:predMSE-GQRM} (for GQRM).

A similar story as before arises out of these plots. For SQRM, note that the prediction RMSE decreases as the number of replicates per time point is increasing and the prediction accuracy is less for higher quantiles. Here, the three methods perform similarly (RMSE ranges between 0.1 and 0.4) for all steps ahead. On the contrary, when the data are generated from any of the other three DGPs, our approach shows much better results than KB and JS. Prediction RMSE for all cases are less than 0.5 for the NP method whereas in KB and JS, the numbers are more than double of that. It can also be observed that, for all of the first three DGPs, there is a discernible difference between the performances of KB and JS methods in higher quantiles. Usually, JS performs better for $\tau=0.95$. For GQRM though, this difference is not present. On the other hand, the nonparametric approach consistently beat the two competing methods and shows less prediction error in all steps ahead in all DGPs. This corroborates the findings of the main paper that any deviation from the assumptions of SQRM affects the performance of the two standard approaches.

\end{document}